\documentclass[runningheads]{llncs}

\usepackage{mathrsfs}
\usepackage{amsmath}
\usepackage{amssymb}

\usepackage{color}
\usepackage{xcolor}

\usepackage{style}

\usepackage{subcaption}
\usepackage{tikz}
\usetikzlibrary{shapes, chains, scopes, positioning,automata}

\usepackage{graphicx}
\usepackage{hyperref}

\usepackage{url}
\newcommand{\doi}[1]{\textsc{doi}: \href{http://dx.doi.org/#1}{\nolinkurl{#1}}}

\usepackage[stable]{footmisc}

\usepackage{enumitem}

\begin{document}
\title{Model Checking Branching Properties on\\ Petri Nets with Transits (Full Version)%
\footnote[3]{This is an extended version of \cite{atva20}.}%
\thanks{This work has been supported by the German Research Foundation (DFG) through Grant Petri Games (392735815) and through the Collaborative Research Center “Foundations of Perspicuous Software Systems” (TRR 248, 389792660), and by the European Research Council (ERC) through Grant OSARES (683300).}%
}
\titlerunning{Model Checking Branching Properties on Petri Nets with Transits}
\author{
Bernd Finkbeiner\inst{1} \and
Manuel Gieseking\inst{2} \and\\
Jesko Hecking-Harbusch\inst{1} \and
Ernst-R\"udiger Olderog\inst{2}
}

\authorrunning{
B.\ Finkbeiner et al.
} 
\institute{
CISPA Helmholtz Center for Information Security, Saarbr\"ucken, Germany\\
\email{\{finkbeiner,jesko.hecking-harbusch\}@cispa.saarland} 
\and
University of Oldenburg, Oldenburg, Germany\\
\email{\{gieseking,olderog\}@informatik.uni-oldenburg.de}
}

\maketitle

\begin{abstract}
To model check concurrent systems, it is convenient to distinguish between the data flow and the control. 
Correctness is specified on the level of data flow whereas the system is configured on the level of control. 
Petri nets with transits and Flow-LTL are a corresponding formalism. 
In Flow-LTL, both the correctness of the data flow and assumptions on fairness and maximality for the control are expressed in linear time.
So far, branching behavior cannot be specified for Petri nets with transits. 
In this paper, we introduce Flow-\ctlStarText{} to express the intended branching behavior of the data flow while maintaining LTL for fairness and maximality assumptions on the control. 
We encode physical access control with policy updates as Petri nets with transits and give standard requirements in Flow-\ctlStarText{}. 
For model checking, we reduce the model checking problem of Petri nets with transits against Flow-\ctlStarText{} via automata constructions to the model checking problem of Petri nets against LTL. 
Thereby, physical access control with policy updates under fairness assumptions for an unbounded number of people can be verified.
\end{abstract}

\section{Introduction}
\label{sec:intro}
\emph{Petri nets with transits}~\cite{DBLP:conf/atva/FinkbeinerGHO19} superimpose a transit relation onto the flow relation of Petri nets. 
The flow relation models the \emph{control} in the form of tokens moving through the net. 
The transit relation models the \emph{data flow} in the form of flow chains.
The configuration of the system takes place on the level of the control whereas correctness is specified on the level of the data flow. 
Thus, Petri nets with transits allow for an elegant separation of the data flow and the control
without the complexity of unbounded colored Petri nets \cite{jensen92}.
We use \emph{physical access control}~\cite{DBLP:conf/vmcai/FrohardtCS11,DBLP:conf/crisis/FitzgeraldTFO12,DBLP:conf/est/GeepallaBD13} as an application throughout the paper. 
It defines and enforces access policies in physical spaces. 
People are represented as the data flow in the building. 
The control defines which policy enforcement points like doors are open to which people identified by their RFID cards~\cite{DBLP:journals/internet/WelbourneBCGRRBB09}. 
Changing access policies is error-prone as closing one door for certain people could be circumvented by an alternative path. 
Therefore, we need to verify such updates. 

\emph{Flow-LTL}~\cite{DBLP:conf/atva/FinkbeinerGHO19} is a logic for Petri nets with transits. 
It specifies linear time requirements on both the control and the data flow. 
Fairness and maximality assumptions on the movement of tokens are expressed in the control part. 
The logic lacks branching requirements for the data flow.
In physical access control, branching requirements can specify that a person has the \emph{possibility} to reach a room but not necessarily has to visit it. 
In this paper, we introduce \emph{Flow-\ctlStarText{}} which maintains LTL to specify the control and adds \ctlStarText{} to specify the data flow. 
Fairness and maximality assumptions in the control part dictate which executions, represented by runs, are checked against the data flow part.

This leads to an interesting encoding for physical access control in Petri nets with transits.
Places represent rooms to collect the data flow. 
Transitions represent doors between rooms to continue the data flow. 
The selection of runs by fairness and maximality assumptions on the control restricts the branching behavior to transitions.
Hence, the data flow is split at transitions: Every room has exactly one outgoing transition enabled unless all outgoing doors are closed. 
This transition splits the data flow into all successor rooms and thereby represents the maximal branching behavior.  

We present a reduction of the model checking problem of safe Petri nets with transits against Flow-\ctlStarText{} to the model checking problem of safe Petri nets against LTL. 
This enables for the first time the automatic verification of physical access control with \emph{policy updates} under fairness and maximality assumptions for an unbounded number of people.
Policy updates occur for example in the evening when every employee is expected to eventually leave the building and therefore access is more restricted.
Such a policy update should prevent people from entering the building but should not trap anybody in the building. 

Our reduction consists of three steps:
First, each data flow subformula of the given Flow-\ctlStarText{} formula is represented, via an alternating tree automaton, an alternating word automaton, and a nondeterministic B\"uchi automaton, by a finite Petri net to guess and then to verify a counterexample tree. 
Second, the original net for the control subformula of the Flow-\ctlStarText{} formula and the nets for the data flow subformulas are connected in sequence. 
Third, an LTL formula encodes the control subformula, the acceptance conditions of the nets for the data flow subformulas, and the correct skipping of subnets in the sequential order.  
This results in a model checking problem of safe Petri nets against LTL.

The remainder of this paper is structured as follows: 
In \refSection{motivation}, we motivate our approach with an example. 
In \refSection{petri-nets}, we recall Petri nets and their extension to Petri nets with transits. 
In \refSection{FlowCTLStar}, we introduce Flow-\ctlStarText{}. 
In \refSection{applications}, we express fairness, maximality, and standard properties for physical access control in Flow-\ctlStarText{}. 
In \refSection{mcFlowCTLStar}, we reduce the model checking problem of Petri nets with transits against Flow-\ctlStarText{} to the model checking problem of Petri nets against LTL.
Section~\ref{sec:related-work} presents related work and \refSection{conclusion} concludes the paper.

\section{Motivating Example}
\label{sec:motivation}
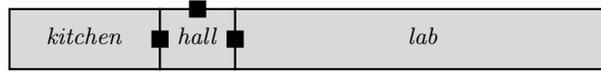
\begin{figure}[t]
\centering
	\begin{tikzpicture}
		\path[use as bounding box] (0,0) rectangle (8, 0.8);
		\draw [-, rectangle, black, fill=black!15] (0, 0) -- (2, 0) -- (2, 0.8) -- (0, 0.8) -- cycle;
		\draw [-, rectangle, black, fill=black!15] (2, 0) -- (2, 0.8) -- (3, 0.8) -- (3, 0) -- cycle;
		\draw [-, rectangle, black, fill=black!15] (3, 0) -- (3, 0.8) -- (8, 0.8) -- (8, 0) -- cycle;
		\node [envplace, opacity = 0] (s0) at (1, -0.1) [label=\emph{kitchen}] {};
		\node [envplace, opacity = 0] (s0) at (2.5, -0.1) [label=\emph{hall}] {};
		\node [envplace, opacity = 0] (s0) at (5.5, -0.1) [label=\emph{lab}] {};
		\draw [-, rectangle, black, fill=black] (1.9, 0.3) -- (2.1, 0.3) -- (2.1, 0.5) -- (1.9, 0.5) -- cycle;
		\draw [-, rectangle, black, fill=black] (2.9, 0.3) -- (3.1, 0.3) -- (3.1, 0.5) -- (2.9, 0.5) -- cycle;
		\draw [-, rectangle, black, fill=black] (2.4, 0.7) -- (2.6, 0.7) -- (2.6, 0.9) -- (2.4, 0.9) -- cycle;
	\end{tikzpicture}
	\caption{
	The layout of a simple building is shown. 
	There are three rooms indicated by gray boxes which are connected by doors indicated by small black boxes.
	}
	\label{fig:layout}
\end{figure}

We motivate our approach with a typical example for physical access control. 
Consider the very simple building layout in \refFig{layout}. 
There are three rooms connected by two doors. 
An additional door is used to enter the building from the outside. 
Only employees have access to the building. 
A typical specification requires that employees can access the \emph{lab} around the clock while allowing access to the \emph{kitchen} only during daytime to discourage too long working hours. 
Meanwhile, certain safety requirements have to be fulfilled like not trapping anybody in the building. 
During the day, a correct access policy allows access to all rooms whereas, during the night, it only allows access to the \emph{hall} and to the \emph{lab}. 

\begin{figure}[t]
\centering
	\begin{tikzpicture}[node distance=1.25cm, on grid, label distance=-1mm]%
		\node [envplace, tokens=1] (s2)  [label=left:{\tiny\emph{hall}}] {};
		
		\node [transition] 	(t0)  [left of = s2, above of = s2, label=left:{\tiny\emph{enterHall}}] {};
		\node [transition] 	(t1)  [left of = s2, below of = s2, label=left:{\tiny\emph{leaveHall}}] {};	
		\draw[thick, shorten >=1pt,<->,shorten <=1pt] (s2) -- (t0);
		\draw[thick, shorten >=1pt,<->,shorten <=1pt] (s2) -- (t1);
		\draw[thick, ->, dashed, cdc_Green, shorten >=1pt, shorten <=1pt] ([xshift=0.1cm, yshift=-0.1cm]t0.east) -- ([xshift=-0.1cm]s2.north);
		\draw[thick,<->, dotted, cdc_Blue, shorten >=1pt, shorten <=1pt] ([xshift=0.1cm, yshift=-0.1cm]t0.south) -- ([yshift=0.1cm]s2.west);
		
		\node [envplace, tokens=1] (s3)  [right of = s2, right of = s2, right of = s2, right of = s2, above of = s2, above of = s2, yshift=-0mm,label=right:{\tiny\emph{lab}}] {};
		\node [envplace, tokens=1] (s4)  [right of = s2, right of = s2, right of = s2, right of = s2, below of = s2, below of = s2, label=right:{\tiny\emph{kitchen}}] {};
		\node [envplace, tokens=1] (s5)  [right of = s2, right of = s2, right of = s2, right of = s2, above of = s2, xshift=-0.625cm, yshift=-0.625cm, label=above:{\tiny~$o_{h\rightarrow l}$}] {};
		\node [envplace] (s6)  [right of = s5, right of = s5, above of = s5, xshift=0.625cm, yshift=-0.625cm, label=above:{\tiny$c_{h\rightarrow l}$}] {};
		\node [envplace, tokens=1] (s7)  [right of = s2, right of = s2, right of = s2, right of = s2, below of = s2, xshift=-0.625cm, yshift=0.625cm, label=above:{\tiny~$o_{h\rightarrow k}$}] {};
		\node [envplace] (s8)  [right of = s7, right of = s7, below of = s7, xshift=0.625cm, yshift=0.625cm, label=above:{\tiny$c_{h\rightarrow k}$}] {};
		
		\node [transition] 	(t11)  [below of = s7, xshift=0.625cm, yshift=0.3125cm, label=245:{\tiny\emph{evening}}] {};
		\draw[thick, shorten >=1pt, ->,shorten <=1pt] (s7) -- (t11);
		\draw[thick, shorten >=1pt, ->,shorten <=1pt] (t11) -- (s8);
		\draw[thick, shorten >=1pt,<->,shorten <=1pt] (t11) -- (s2);
		\draw[thick, shorten >=1pt,<->,shorten <=1pt] (t11) -- (s3);
		\draw[thick, shorten >=1pt,<->,shorten <=1pt] (t11) -- (s4);
		\draw[thick, <->, dashed, cdc_Green, shorten >=1pt, shorten <=1pt] ([xshift=0.3cm, yshift=-0.1cm]s2.east) -- ([xshift=-0.1cm, yshift=0.2cm]t11.west);
		\draw[thick, <->, dashdotted, orange, shorten >=1pt, shorten <=1pt] ([xshift=0.1cm, yshift=-0.1cm]s3.south) -- ([xshift=0.1cm, yshift=0.1cm]t11.north);
		\draw[thick, <->, dotted, cdc_Blue, shorten >=1pt, shorten <=1pt] ([xshift=0.2cm]s4.north) -- ([xshift=0.2cm]t11.south);
		
		\node [transition] 	(t2)  [right of = s2, right of = s2, right of = s2, right of = s2, right of = s2, right of = s2, label=right:{\tiny\emph{hall}$\rightarrow$[\emph{l},\emph{k}]}] {};
		\draw[thick, shorten >=1pt,<->,shorten <=1pt] (s2) -- (t2);
		\draw[thick, shorten >=1pt,<->,shorten <=1pt, dashdotted, orange] (s3) -- (t2);
		\draw[thick, shorten >=1pt,<->,shorten <=1pt, dotted, cdc_Blue] (s4) -- (t2);
		\draw[thick, shorten >=1pt,<->,shorten <=1pt] (s5) -- (t2);
		\draw[thick, shorten >=1pt,<->,shorten <=1pt] (s7) -- (t2);
		
		\draw[thick, ->, dashed, cdc_Green, shorten >=1pt, shorten <=1pt] ([yshift=0.1cm]s2.east) -- ([xshift=-0.1cm, yshift=0.1cm]t2.west);
		\draw[thick, ->, dashed, cdc_Green, shorten >=1pt, shorten <=1pt] ([yshift=0.2cm]t2.north) -- ([xshift=0.2cm, yshift=-0.1cm]s3.east);
		\draw[thick,<->, shorten >=1pt, shorten <=1pt] ([xshift=-0.1cm, yshift=0.1cm]t2.north) -- ([yshift=-0.1cm]s3.east);
		\draw[thick, ->, dashed, cdc_Green, shorten >=1pt, shorten <=1pt] ([yshift=-0.2cm]t2.south) -- ([xshift=0.2cm, yshift=0.1cm]s4.east);
		\draw[thick,<->, shorten >=1pt, shorten <=1pt] ([xshift=-0.1cm, yshift=-0.1cm]t2.south) -- ([yshift=0.1cm]s4.east);
		
		\node [transition] 	(t3)  [right of = s2, above of = s2, yshift=0mm, label=above:{\tiny\emph{hall}$\rightarrow$\emph{lab}~~~~~~}] {};
		\draw[thick, shorten >=1pt,<->,shorten <=1pt] (s2) -- (t3);
		\draw[thick, shorten >=1pt,<->,shorten <=1pt] (s3) -- (t3);
		\draw[thick, shorten >=1pt,<->,shorten <=1pt] (s5) -- (t3);
		\draw[thick, shorten >=1pt,<->,shorten <=1pt] (s8.140) -- (t3);
		\node [transition] 	(t4)  [right of = s2, below of = s2, label=below:{\tiny\emph{hall}$\rightarrow$\emph{kitchen}~~~~~~~~~~}] {};
		\draw[thick, shorten >=1pt,<->,shorten <=1pt] (s2) -- (t4);
		\draw[thick, shorten >=1pt,<->,shorten <=1pt] (s4) -- (t4);
		\draw[thick, shorten >=1pt,<->,shorten <=1pt] (s6) -- (t4);
		\draw[thick, shorten >=1pt,<->,shorten <=1pt] (s7) -- (t4);
		\node [transition] 	(t5)  [left of = t3, above of = t3, yshift=-0mm, label=left:{\tiny\emph{lab}$\rightarrow$\emph{hall}}] {};
		\draw[thick, shorten >=1pt,<->,shorten <=1pt] (s2) -- (t5);
		\draw[thick, shorten >=1pt,<->,shorten <=1pt] (s3) -- (t5);
		\node [transition] 	(t6)  [left of = t4, below of = t4, label=left:{\tiny\emph{kitchen}$\rightarrow$\emph{hall}}] {};
		\draw[thick, shorten >=1pt,<->,shorten <=1pt] (s2) -- (t6);
		\draw[thick, shorten >=1pt,<->,shorten <=1pt] (s4) -- (t6);
		
		\draw[thick, ->, dashed, cdc_Green, shorten >=1pt, shorten <=1pt] ([xshift=0.2cm, yshift=0.1cm]s2.north) -- ([xshift=-0.1cm, yshift=-0.1cm]t3.west);
		\draw[thick, ->, dashed, cdc_Green, shorten >=1pt, shorten <=1pt] ([xshift=0.2cm, yshift=-0.1cm]s2.south) -- ([xshift=-0.1cm, yshift=0.1cm]t4.west);
		\draw[thick,<->, dotted, cdc_Blue, shorten >=3pt, shorten <=5pt] (t3.east) -- ([yshift=-0.2cm]s3.west);
		\draw[thick, ->, dashed, cdc_Green, shorten >=6pt, shorten <=3pt] ([yshift=0.2cm]t3.east) -- (s3.west);
		\draw[thick,<->, dotted, cdc_Blue, shorten >=3pt, shorten <=5pt] (t4.east) -- ([yshift=0.2cm]s4.west);
		\draw[thick, ->, dashed, cdc_Green, shorten >=6pt, shorten <=3pt] ([yshift=-0.2cm]t4.east) -- (s4.west);
		
		\draw[thick, ->, dashed, cdc_Green, shorten >=5pt, shorten <=5pt] ([yshift=0.1cm]s3.west) -- ([yshift=0.1cm]t5.east);
		\draw[thick, ->, dashed, cdc_Green, shorten >=5pt, shorten <=5pt] ([yshift=-0.1cm]s4.west) -- ([yshift=-0.1cm]t6.east);
		
		\draw[thick, <-, dashed, cdc_Green, shorten >=1pt, shorten <=2pt] ([xshift=-0.1cm, yshift=0.2cm]s2.north) -- ([xshift=-0.1cm, yshift=-0.1cm]t5.south);
		\draw[thick,<->, dotted, cdc_Blue, shorten >=1pt, shorten <=2pt] ([xshift=0.1cm, yshift=0.2cm]s2.north) -- ([xshift=0.1cm, yshift=-0.1cm]t5.south);
		\draw[thick, <-, dashed, cdc_Green, shorten >=1pt, shorten <=2pt] ([xshift=-0.1cm, yshift=-0.2cm]s2.south) -- ([xshift=-0.1cm, yshift=0.1cm]t6.north);
		\draw[thick,<->, dotted, cdc_Blue, shorten >=1pt, shorten <=2pt] ([xshift=0.1cm, yshift=-0.2cm]s2.south) -- ([xshift=0.1cm, yshift=0.1cm]t6.north);
	\end{tikzpicture}
	\caption{The Petri net with transits encoding the building from \refFig{layout} is depicted. Rooms are modeled by corresponding places, doors by transitions. Tokens in places starting with $o$ configure the most permissive access policy during the day. In the \emph{evening}, access to the \emph{kitchen} is restricted. Employees in the building are modeled by the transit relation depicted by green, blue, and orange arrows. 
	}
	\label{fig:motivation}
\end{figure}
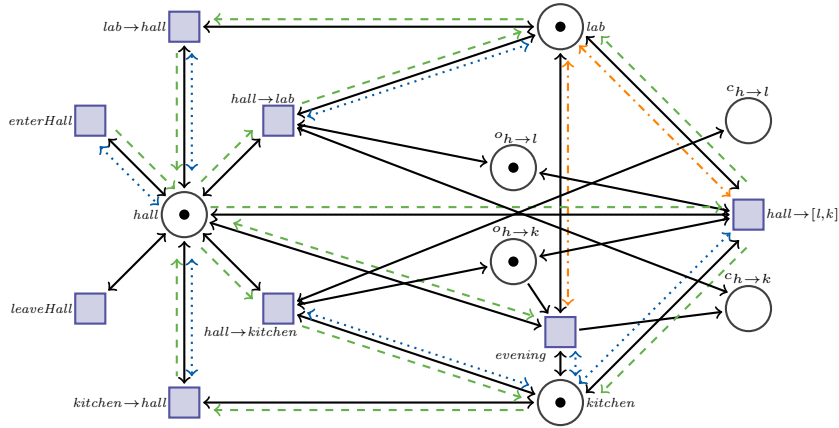

Figure~\ref{fig:motivation} shows a Petri nets with transits modeling the building layout from \refFig{layout}. 
There are corresponding places (represented by circles) with tokens (represented by dots) for the three rooms: \emph{hall}, \emph{lab}, and \emph{kitchen}. 
These places are connected by transitions (represented by squares) of the form \emph{from}$\rightarrow$\emph{to} for \emph{from} and \emph{to} being rooms. 
The doors from the \emph{kitchen} and \emph{lab} to the \emph{hall} cannot be closed as this could trap people. 
For all other doors, places of the form $o_{\textit{from}\rightarrow\textit{to}}$ and $c_{\textit{from}\rightarrow\textit{to}}$ exist to represent whether the door is open or closed. 

In (safe) Petri nets, transitions define the movement of tokens: Firing a transition removes one token from each place with a black arrow leaving to the transition and adds one token to each place with a black arrow coming from the transition. 
Firing transition \emph{evening} moves one token from place $o_{h\rightarrow k}$ to place~$c_{h\rightarrow k}$ as indicated by the single-headed, black arrows and one token from and to each of the places \emph{hall}, \emph{lab}, and \emph{kitchen} as indicated by the double-headed, black arrows.
Firing transitions modeling doors returns all tokens to the same places while the transit relation as indicated by the green, blue, and orange arrows represents employees moving through the building. 
Dashed and dotted arrows only distinguish them from black arrows in case colors are unavailable. 
 
Firing transition \emph{enterHall} starts a flow chain modeling an employee entering the building as indicated by the single-headed, green (dashed) arrow. 
Meanwhile, the double-headed, blue (dotted) arrow maintains all flow chains previously in \emph{hall}.
All flow chains collectively represent the data flow in the modeled system incorporating all possible control changes.
Firing transitions \emph{from}$\rightarrow$\emph{to}, which correspond to doors, continues all flow chains from place \emph{from} to place \emph{to} as indicated by the single-headed, green (dashed) arrows and merges them with all flow chains in the place \emph{to} as indicated by the double-headed, blue (dotted) arrows. 
For example, firing transition \emph{hall}$\rightarrow$\emph{lab} lets all employees in the \emph{hall} enter the \emph{lab}. 
When employees \emph{leave} the \emph{hall}, their flow chain ends because it is not continued as indicated by the lack of colored arrows at transition \emph{leaveHall}. 

Flow-\ctlStarText{} allows the splitting of flow chains in transitions. 
Splitting flow chains corresponds to branching behavior. 
Thus, when the doors to the \emph{lab} and \emph{kitchen} are open, we represent this situation by \emph{one} transition which splits the flow chains. 
Transition \emph{hall}$\rightarrow$[\emph{l},\emph{k}] realizes this by the single-headed, green (dashed) arrows from the \emph{hall} to the \emph{lab} and \emph{kitchen}. 
Branching results in a \emph{flow tree} for the possible behavior of an employee whereas a flow chain represents one explicit path from this flow tree, i.e., each employee has one flow tree with possibly many flow chains. 
Notice that transition \emph{hall}$\rightarrow$[\emph{l},\emph{k}] can only be fired during the day, because, when firing transition \emph{evening}, access to the \emph{kitchen} is revoked. 
Then, only transition \emph{hall}$\rightarrow$\emph{lab} can be fired for moving flow chains from the \emph{hall}.
For simplicity, we restrict the example to only one time change which implies that the transition \emph{hall}$\rightarrow$\emph{kitchen} can never be fired. 
Firing transition \emph{evening} continues all flow chains in the three places \emph{hall}, \emph{lab}, and \emph{kitchen}, respectively, as indicated by the distinctly colored, double-headed arrows. 
Thus, we can specify requirements for the flow chains after the time change. 

We specify the correctness of access policies with formulas of the logic \emph{Flow-\ctlStarText{}}. 
The formula $\A\,\ctlAG \ctlEF \textit{lab}$ expresses \emph{persistent permission} requiring that all flow chains ($\A$) on all paths globally ($\ctlAG$) have the possibility~($\ctlEF$) to reach the \emph{lab}. 
The formula $\A\,\ctlAU{(\ctlEF \textit{kitchen})}{\textit{evening}} $ expresses \emph{dependent permission} requiring that all flow chains on all paths~($\ctlAll$) have the possibility to reach the \emph{kitchen} until ($\ctlUntil$) \emph{evening}. 
Both properties require weak or strong fairness for all transitions modeling doors to be satisfied. 
The second property additionally requires weak or strong fairness for transition \emph{evening} to be satisfied. 
Flow-\ctlStarText{} and specifying properties with it are discussed further in \refSection{FlowCTLStar} and \refSection{applications}.

\section{Petri Nets with Transits}
\label{sec:petri-nets}
We recall the formal definition of {Petri nets with transits}~\cite{DBLP:conf/atva/FinkbeinerGHO19} as extension of Petri nets~\cite{DBLP:books/sp/Reisig85a}.
We refer the reader to Appendix~\ref{app:pn} for more details.
A safe \emph{Petri net} is a structure \(\petriNet\) with the set of \emph{places}~$\pl$, the set of \emph{transitions}~$\tr$, 
the (\emph{control}) \emph{flow relation}~$\fl \subseteq (\pl \times \tr) \cup (\tr \times \pl)$, and the \textit{initial marking}~$\init\subseteq\pl$.
In \emph{safe} Petri nets, each reachable marking contains at most one token per place. 
The elements of the disjoint union $\pl\cup\tr$ are considered as \textit{nodes}. 
We define the \textit{preset} (and \textit{postset}) of a node~$x$ from Petri net~$\pNet$ as $\pre{\pNet}{x} = \{y \in \pl \cup \tr \mid (y,x)\in\fl\}$ (and $\post{\pNet}{x} = \{y \in \pl \cup \tr \mid (x,y)\in\fl\}$). 
A safe \emph{Petri net with transits} is a structure \(\petriNetFl\)
which additionally contains a \emph{transit relation}~$\tokenflow$ refining the flow relation of the net to define the data flow.
For each transition~$t \in \transitions$, $\tokenflow(t)$ is a relation of type $\tokenflow(t) \subseteq (\pre{\pNet}{t}\cup\{\startfl\}) \times \post{\pNet}{t}$, where the symbol~$\startfl$ denotes a \emph{start}. 
With $\startfl\ \tokenflow(t)\ q$, we define the start of a new data flow in place \(q\) via transition~$t$
and with $p\ \tokenflow(t)\ q$ that all data in place~$p$ \emph{transits} via transition~$t$ to place~$q$. 
The \emph{postset regarding \(\tokenflow\)} of a place \(p\in\pl\) and a transition \(t\in\post{\pNet}{p}\) is defined by
\(\post{\tokenflow}{p,t}=\{p'\in\pl\with (p,p')\in\tokenflow(t)\}\).

The graphic representation of $\tokenflow(t)$ 
in Petri nets with transits uses a \emph{color coding} as can be seen in 
\refFig{motivation}.
Black arrows represent the usual \emph{control flow}.
Other matching colors per transition are used to represent the transits of the \emph{data flow}.
Transits allow us to specify where the
data flow is moved forward, split, and merged, where it ends,
and where data is newly created.
The data flow can be of infinite length
and at any point in time (possibly restricted by the control)
new data can enter the system at different locations.

As the data flow is a local property of each distributed
component (possibly shared via joint transitions)
it is convenient that Petri nets with transits
use a true concurrency semantics
to define the data flow.
Therefore, we recall the notions of unfoldings and runs~\cite{DBLP:journals/acta/Engelfriet91,DBLP:series/eatcs/EsparzaH08}
and their application to Petri nets with transits. 
In the unfolding of a Petri net $\pNet$, every transition stands for the unique occurrence
(instance) of a transition of $\pNet$ during an execution.
To this end, every loop in~$\pNet$ is unrolled and
every backward branching place 
is expanded by multiplying the place.
Forward branching, however, is preserved. Formally, an \emph{unfolding}
is a branching process $\beta^U = (\pNet^U, \lambda^U)$ 
consisting of an occurrence net~$\pNet^U$ and a
homomorphism $\lambda^U$ that labels the places and transitions in $\pNet^U$
with the corresponding elements of $\pNet$.
The unfolding exhibits concurrency, causality, and nondeterminism (forward branching)
of the unique occurrences of the transitions in $\pNet$ during all possible
executions.
A \emph{run} of $\pNet$ is a subprocess~$\runPN = (\pNet^R, \rho)$ of~$\beta^U$,
where \(\forall p\in\pl^R:|\post{\pNet^R}{p}|\leq 1\) holds,
i.e., all nondeterminism has been resolved but concurrency is preserved.
Thus, a run formalizes one concurrent execution of $\pNet$.
We lift the transit relation of a Petri net with transits
to any branching process and thereby obtain notions of runs
and unfoldings for Petri nets with transits.
Consider a run $\runPN = (\mathcal N^R, \rho)$ of $\pNet$ and 
a finite or infinite firing sequence 
$\zeta = M_0\firable{t_0}M_1\firable{t_1}M_2\cdots $ of $\pNet^R$ with 
$M_0 = \initialmarking^{R}$. This sequence \emph{covers}~$\runPN$ if 
\(
  (\forall p \in \places^{R} : \exists i \in \mathbb{N}: p \in M_i)
  \land
  (\forall t \in \transitions^{R} : \exists i \in \mathbb{N}: t = t_i),
\)
i.e., all places and transitions in $\mathcal N^R$ appear in $\zeta$.
Several firing sequences may cover~$\runPN$.

We define flow chains by following the transits of a given run.
A (\emph{data}) \emph{flow chain}
of a run \(\runPN=(\pNet^R,\rho)\) of a Petri net with transits \(\pNet\)
is a \emph{maximal} sequence \(\flowChain =  t_0, p_0, t_1, p_1, t_2 \dots\)
of connected places and transitions of \(\pNet^R\) with
\begin{enumerate}[topsep=6pt]
	\item[(I)] \((\startfl,p_0)\in\tfl^R(t_0)\),	
	\item[(con)] \((p_{i-1}, p_{i})\in\tfl^R(t_i)\) for all \(i\in\N\setminus\{0\}\) 
					if \(\flowChain\) is infinite and for all \(i\in\{1,\ldots n\}\)
					if \(\flowChain=t_0, p_0, t_1,\ldots, t_{n}, p_n\) is finite,  
	\item[(max)] if \(\flowChain=t_0, p_0, t_1,\ldots, t_n, p_n\) is finite
	      there is no transition \(t\in\tr^R\) and place \(q\in\pl^R\) such that \((p_n,q)\in\tfl^R(t)\).
\end{enumerate}
A \(\emph{flow chain suffix}\) \(\flowChain' =  t_0, p_0, t_1, p_1, t_2 \dots\) of a run \(\runPN\)
requires constraints (con), (max), and in addition to (I) allows 
that the chain has already started, i.e.,
\(\exists p\in\pl^R: (p,p_0)\in\tfl^R(t_0)\).

\begin{figure}[t]
\centering
\begin{tikzpicture}[node distance=1cm, on grid, label distance=-1mm]%
\tikzset{
		flowA/.style ={flow={dotted,cdc_Blue}{}},
		flowB/.style ={flow={dashed,cdc_Green}{}},
		flowC/.style ={flow={dashdotted,orange}{}},
}
	\node [envplace, tokens=1] 			(h0)  [label=above:{\tiny\(\mathit{hall}_0\)}] {};
	\node [transition, below=of h0] 	(enter)  [label=below:{\tiny\(\mathit{enterHall}_0\)}] {};
	\node [envplace, right=of enter] 	(h1)  [label=above:{\tiny\(\mathit{hall}_1\)}] {};
	\node [transition, right=of h1] 	(tlk) {};
	\node [envplace, right=of tlk] 		(h2)  [label=below:{\tiny\(\mathit{hall}_2\)}] {};
	\node [transition, right=of h2] 	(enter2)  [label=below:{\tiny\(\mathit{enterHall}_1\)}] {};
	\node [envplace, right=of enter2] 	(h3)  [label=below:{\tiny\(\mathit{hall}_3\)}] {};
	\node [transition, right=of h3] 	(hl)  [label=below:{\tiny\emph{lab}$\rightarrow$\emph{hall}}] {};
	\node [envplace, right=of hl] 		(h4)  [label=above:{\tiny\(\mathit{hall}_4\)}] {};
	\node [transition, right=of h4] 	(hk)  [label=above:{\tiny\emph{kitchen}$\rightarrow$\emph{hall}}] {};
	\node [envplace, right=of hk] 		(h5)  [label=above:{\tiny\(\mathit{hall}_5\)}] {};
	\node [transition, right=of h5] 	(eve) [label={[xshift=-1mm]below right:{\tiny\(\mathit{evening}\)}}] {};
	\node [envplace, right=of eve] 		(h6)  [label=above:{\tiny\(\mathit{hall}_6\)}] {};

	\node [envplace, tokens=1,above=of h1,xshift=-3.5mm]		(ohl) [label=above:{\tiny$o_{h\rightarrow l}$}] {};
	\node [envplace, right=of ohl,xshift=-3.5mm]				(ohl1) [label=above:{\tiny$o_{h\rightarrow l}'$}] {};
	\node [envplace, tokens=1,above=of tlk] 	(l0) [label=above:{\tiny\(\mathit{lab}_0\)}] {};
	\node [envplace, right=of l0]				(l1) [label=above:{\tiny\(\mathit{lab}_1\)}] {};
	\node [envplace, tokens=1,below=of enter]	(ohk) [label=below:{\tiny$o_{h\rightarrow k}$}] {};
	\node [envplace, right=of ohk]				(ohk1) [label=below:{\tiny$o_{h\rightarrow k}'$}] {};
	\node [envplace, tokens=1,right=of ohk1] 	(k0) [label=below:{\tiny\(\mathit{kitchen}_0\)}] {};
	\node [envplace, right=of k0]				(k1) [label=below:{\tiny\(\mathit{kitchen}_1\)}] {};

	\node [envplace, above=of hl]				(l2) [label=above:{\tiny\(\mathit{lab}_2\)}] {};
	\node [envplace, below=of hk]				(k2) [label=below:{\tiny\(\mathit{kitchen}_2\)}] {};

	\node [envplace, above=of eve]				(l3) [label=above:{\tiny\(\mathit{lab}_3\)}] {};
	\node [envplace, below=of eve]				(k3) [label=below:{\tiny\(\mathit{kitchen}_3\)}] {};
	\node [envplace, above=of h6] 				(chk)[label=above:{\tiny$c_{h\rightarrow k}$}] {};

		\addFlowAdditionalArrows{enter}{h0}{flowA}{pre}
		\addFlowAdditionalArrows{enter}{h1}{flowB,flowA}{post}

		\addFlowAdditionalArrows{h1}{tlk}{flowB}{post}
		\addFlowAdditionalArrows{tlk}{h2}{}{post}
		\addFlowAdditionalArrows{tlk}{l0}{flowC}{pre}
		\addFlowAdditionalArrows{k0}{tlk}{flowA}{post}
		\addFlowAdditionalArrows{tlk}{l1}{flowB,flowC}{post}
		\addFlowAdditionalArrows{tlk}{k1}{flowA,flowB}{post}
		\addFlowAdditionalArrows{tlk}{ohl1}{}{post}
		\addFlowAdditionalArrows{tlk}{ohl}{}{pre}
		\addFlowAdditionalArrows{tlk}{ohk1}{}{post}
		\addFlowAdditionalArrows{tlk}{ohk}{}{pre}

		\addFlowAdditionalArrows{enter2}{h2}{flowA}{pre}
		\addFlowAdditionalArrows{enter2}{h3}{flowB,flowA}{post}

		\addFlowAdditionalArrows{hl}{h3}{flowA}{pre}
		\addFlowAdditionalArrows{l1}{hl}{flowB}{post,bend left=10}
		\addFlowAdditionalArrows{hl}{h4}{flowB,flowA}{post}
		\addFlowAdditionalArrows{l2}{hl}{}{pre}

		\addFlowAdditionalArrows{hk}{h4}{flowA}{pre}
		\addFlowAdditionalArrows{hk}{k1}{flowB}{pre,bend left=20}
		\addFlowAdditionalArrows{hk}{h5}{flowB,flowA}{post}
		\addFlowAdditionalArrows{k2}{hk}{}{pre}

		\addFlowAdditionalArrows{h5}{eve}{flowB}{post}
		\addFlowAdditionalArrows{l2}{eve}{flowC}{post, bend left=20}
		\addFlowAdditionalArrows{eve}{k2}{flowA}{pre}
		\addFlowAdditionalArrows{l3}{eve}{flowC}{pre}
		\addFlowAdditionalArrows{eve}{k3}{flowA}{post}
		\addFlowAdditionalArrows{eve}{h6}{flowB}{post}
		\addFlowAdditionalArrows{chk}{eve}{}{pre}

\draw[-,line width=8mm, gray,opacity=0.2, rounded corners] (enter.west) -- (tlk.center) to[in=182] (l1.center);
\path (l1.center) edge [out=145,-,line width=8mm, gray,opacity=0.2, rounded corners,bend left=20] (hl.center);
\draw[-,line width=8mm, gray,opacity=0.2, rounded corners]  (hl.center) to[out=-38, in=180] (h4) -- (h6.east);
\draw[-,line width=8mm, gray,opacity=0.2, rounded corners]  ([xshift=4.5mm, yshift=-1.1mm]tlk.center) to[out=-46,in=171] (k1.center);
\draw[-,line width=8mm, gray,opacity=0.2, rounded corners]  (k1.center) to[out=-10,in=-90,looseness=0.5] ([xshift=-4mm,yshift=-4mm]hk.center);

\end{tikzpicture}
	\caption{A finite run of the Petri net with transits from \refFig{motivation} with two data flow trees is depicted. The first one is indicated by the gray shaded area. 
	}
	\label{fig:runFlowTrees}
\end{figure}
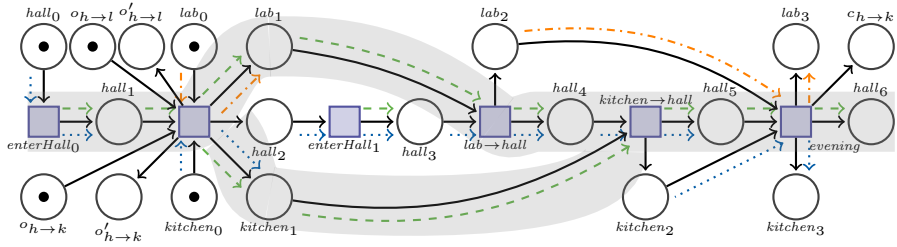

A \(\Sigma\)-labeled \emph{tree} over a set of \emph{directions} \(\D\subset\N\) is
a tuple \((\tree,\treeLab)\), 
with a labeling function \(\treeLab:\tree\to\Sigma\) and 
a tree \(\tree\subseteq\D^*\) such that
if \(x\cdot c\in\tree\) for \(x\in\D^*\) and \(c\in\D\), then 
both \(x\in\tree\) and for all \(0\leq c'<c\) also \(x\cdot c'\in\tree\) holds.
A (\emph{data}) \emph{flow tree} of a run \(\runPN=(\pNet^R,\rho)\)
represents all branching behavior in the transitions of the run
w.r.t.\ the transits.
Formally, for each \(t_0\in\tr^R\) and place \(p_0\in\pl^R\)
with \((\startfl,p_0)\in\tfl^R(t_0)\), 
there is a 
\(\tr^R\times\pl^R\)-labeled tree \(\flowtree=(\tree,\treeLab)\)
over directions
\(\D\subseteq\{0,\ldots,\mathtt{max}\{\size{\post{\tfl^R}{p,t}}-1 \with p\in\pl^R\wedge t\in\post{\pNet^R}{p}\}\}\)
with
\begin{enumerate}[topsep=6pt]
\item \(\treeLab(\epsilon)=(t_0,p_0)\) for the root \(\epsilon\), and
\item if \(n\in\tree\) with \(\treeLab(n)=(t,p)\) then for the only transition \(t'\in\post{\pNet^R}{p}\) (if existent)
we have for all \(0\leq i<\size{\post{\tfl^R}{p,t'}}\) that \(n\cdot i\in\tree\)
with \(\treeLab(n\cdot i)=(t',q)\) for \(q=\langle\post{\tfl^R}{p,t'}\rangle_i\) 
where \(\langle\post{\tfl^R}{p,t'}\rangle_i\) is the \(i\)-th value of the ordered list \(\langle\post{\tfl^R}{p,t'}\rangle\).
\end{enumerate}
Figure~\ref{fig:runFlowTrees} shows a finite run of the example from \refFig{motivation} with two flow trees.
The first tree starts with transition \(\mathit{enterHall}_0\), i.e., \(\treeLab(\epsilon)=(\mathit{enterHall}_0,\mathit{hall}_1)\)
and is indicated by the gray shaded area.
This tree represents an extract of the possibilities of a person entering the hall during the
day ending with the control change to the evening policy.
The second tree
(\(\treeLab(\epsilon)=(\mathit{enterHall}_1,\mathit{hall}_3),\treeLab(0)=(\mathit{lab}\text{$\rightarrow$}\mathit{hall},\mathit{hall}_4),\treeLab(00)=(\mathit{kitchen}\text{$\rightarrow$}\mathit{hall},\mathit{hall}_5),\treeLab(000)=(\mathit{evening},\mathit{hall}_6)\))
shows the possibilities of a person in this run who
later enters the hall and can, because of the run, only stay there.
Note that the trees only end due to the finiteness of the run.
For maximal runs, trees can only end 
when transition \(\mathit{leaveHall}\) is fired.

\section{Flow-\ctlStarText{} for Petri Nets with Transits}
\label{sec:FlowCTLStar}
We define the new logic \emph{Flow-\ctlStarText{}} to reason 
about the Petri net behavior and the data flow individually.
Properties on the selection of runs and the general behavior of the net 
can be stated in \ltlText{}, requirements on the data flow in \ctlStarText{}.
\subsection{\ltlText{} on Petri Net Unfoldings}
\label{sec:ltl}
We recall \ltlText{} with \emph{atomic propositions} \(\AP=\pl\cup\tr\) on a Petri net \(\petriNet\)
and define the semantics on runs and their firing sequences.
We use the \emph{ingoing} semantics, i.e.,
we consider the marking and the transition used to enter the marking,
and \emph{stutter} in the last marking for finite firing sequences.

\emph{Syntactically}, the set of \emph{linear temporal logic}~(LTL) formulas \(\mathtt{LTL}\) over \(\AP\)
is defined by
\(\phiLTL ::= \true \mid a \mid \lnot \phiLTL \mid \phiLTL_1 \land \phiLTL_2 \mid
			 \ltlNext \phiLTL \mid \phiLTL_1 \ltlUntil \phiLTL_2,\)
with \(a\in\AP\) and \(\ltlNext\)~being the \emph{next} and \(\ltlUntil\) the \emph{until} operator.
As usual, we use the propositional operators \(\vee\), \(\rightarrow\), and \(\leftrightarrow\),
the temporal operators \(\ltlEventually \phiLTL=\true\ltlUntil\phiLTL\) (the \emph{eventually} operator)
and \(\ltlAlways\phiLTL=\lnot\ltlEventually\lnot\phiLTL\) (the \emph{always} operator)
as abbreviations.

For a Petri net \(\pNet\), we define a \emph{trace} as a mapping \(\trace:\N\to2^\AP\).
The \(i\)-th suffix \(\trace^i:\N\to2^\AP\) is a trace defined by \(\trace^i(j)=\trace(j+i)\) for all \(j\in\N\).
To a (finite or infinite) covering firing sequence \(\firingSequence= M_0\firable{t_0}M_1\firable{t_1}M_2\cdots\) 
of a run \(\runPN=(\pNet^R,\rho)\) of \(\pNet\), we associate
a trace \(\trace(\firingSequence):\N\to 2^\AP\) with
\(\trace(\firingSequence)(0) = \rho(M_0)\),
\(\trace(\firingSequence)(i) = \{ \rho(t_{i-1})\}\cup\rho(M_i) \) for all \(i\in\N\setminus\{0\}\) if \(\firingSequence\) is infinite
and 
\(\trace(\firingSequence)(i) = \{ \rho(t_{i-1})\}\cup\rho(M_i)\) for all \(0 < i \leq n\), and
\(\trace(\firingSequence)(j) = \rho(M_n)\) for all \(j > n\) if \(\firingSequence=M_0\firable{t_0} \cdots \firable{t_{n-1}}M_n\)
is finite.
Hence, a trace of a firing sequence covering a run is an infinite sequence of states collecting the corresponding marking
and ingoing transition of \(\pNet\), which stutters on the last marking for finite sequences.

The \emph{semantics} of an LTL formula \(\phiLTL \in \mathtt{LTL}\) on a Petri net \(\pNet\)
is defined over the traces of the covering firing sequences of its runs:
\(\pNet \models_\mathtt{LTL}\phiLTL\) iff for all runs \(\runPN\) of \(\pNet:\ \runPN \models_\mathtt{LTL} \phiLTL\),
\(\runPN \models_\mathtt{LTL}\phiLTL\) iff for all firing sequences \(\firingSequence\) covering \(\runPN:\ \sigma(\firingSequence) \models_\mathtt{LTL} \phiLTL\),
\(\sigma \models_\mathtt{LTL}\true\),
\(\sigma \models_\mathtt{LTL}a\) iff \(a \in \sigma(0)\),
\(\sigma \models_\mathtt{LTL}\lnot\phiLTL\) iff not \(\sigma \models_\mathtt{LTL} \phiLTL\),
\(\sigma \models_\mathtt{LTL}{\phiLTL}_1 \land {\phiLTL}_2\) iff \(\sigma \models_\mathtt{LTL} {\phiLTL}_1\) and \(\sigma \models_\mathtt{LTL} {\phiLTL}_2\),
\(\sigma \models_\mathtt{LTL}\ltlNext\phiLTL\) iff \(\sigma^1 \models_\mathtt{LTL} \phiLTL \), and
\(\sigma \models_\mathtt{LTL}{\phiLTL}_1\ltlUntil{\phiLTL}_2\) iff there exists a \(j\ge 0\) with \(\sigma^j \models_\mathtt{LTL} {\phiLTL}_2\) and \(\sigma^i \models_\mathtt{LTL} {\phiLTL}_1\) holds for all \(0 \le i < j\) .

\subsection{\ctlStarText{} on Flow Chains}
\label{sec:ctlStar}
To specify the data flow of a Petri net with transits \(\petriNetFl\), we use the complete \emph{computation tree
logic~(\ctlStarText)}. The set of \ctlStarText{} formulas~\(\mathtt{CTL^*}\) over \(\AP=\pl\cup\tr\)
is given by the following \emph{syntax} of \emph{state formulas}:
\(
 \stateFormula ::= a \mid \lnot\,\stateFormula \mid \stateFormula_1 \land \stateFormula_2 \mid \ctlExists\, \pathFormula
\)
where \(a \in \AP\), \(\stateFormula\), \(\stateFormula_1\), \(\stateFormula_2\) are state formulas,
and \(\pathFormula\) is a \emph{path formula} with the following \emph{syntax}:
\(
 \pathFormula ::= \stateFormula \mid \lnot\,\pathFormula \mid \pathFormula_1 \land \pathFormula_2 \mid \ctlNext\,\pathFormula \mid \pathFormula_1 \ctlUntil\, \pathFormula_2
\)
where~\(\stateFormula\) is a state formula and \(\pathFormula\), \(\pathFormula_1,\), \(\pathFormula_2\) are path formulas.
We use the propositional operators
\(\vee\), \(\rightarrow\), \(\leftrightarrow\),
the path quantifier \(\ctlAll\pathFormula=\lnot\ctlExists\lnot\pathFormula\),
and the temporal operators \(\ctlFinally\pathFormula=\true\,\ctlUntil\,\pathFormula\),
\(\ctlGlobally\pathFormula=\lnot\ctlFinally\lnot\pathFormula\), \(\pathFormula_1\ctlRelease\pathFormula_2=\lnot(\lnot\pathFormula_1\ctlUntil\lnot\pathFormula_2)\)
as abbreviations.

To a (finite or infinite) flow chain suffix \(\flowChain=t_0,p_0,t_1,p_{1},t_{2},\ldots\)
of a run \(\runPN=(\pNet^R,\rho)\) of \(\pNet\), we associate
a trace \(\trace(\flowChain):\N\to\mathcal{S}=\{\{t,p\},\{p\} \with p\in\pl^R\wedge t\in\tr^R\}\)
with \(\trace(\flowChain)(i) = \{t_i,p_i\}\) for all \(i\in\N\) if \(\flowChain\) is infinite
and 
\(\trace(\flowChain)(i) = \{t_i,p_i\}\) for all \(i \leq n\), and
\(\trace(\flowChain)(j) = \{p_n\}\) for all \(j > n\) if \(\flowChain=t_0,p_0,t_1,p_{1},\ldots,t_{n},p_n\)
is finite.
Hence, a trace of a flow chain suffix is an infinite sequence of states collecting
the current place and ingoing transition of the flow chain,
which stutters on the last place \(p\) of a finite flow chain suffix.
We define \(\trace_\stutterSymbol(\{p\})(i)=\{p\}\) for all \(i\in\N\)
to stutter on the last place of a finite flow chain suffix.

The \emph{semantics} of a computation tree logic formula \(\ctlStarFormula\in\mathtt{CTL^*}\) is evaluated 
on a given run \(\runPN=(\pNet^R,\rho)\) of the Petri net with transits \(\pNet\) and a state \(s\in\mathcal{S}\)
of a trace \(\trace(\flowChain)\) of a flow chain suffix \(\flowChain\) or the trace itself:
\[
\begin{array}{lll}
\runPN, s \satCTLStar a 			  			  					&\ \text{iff}\ & 	a \in \rho(s)						 \\[1mm]
\runPN, s \satCTLStar \lnot\, \stateFormula         				&\ \text{iff}\ & 	\text{not}\ \runPN, s \satCTLStar \stateFormula \\[1mm]
\runPN, s \satCTLStar \stateFormula_1 \land \stateFormula_2      &\ \text{iff}\ &    \runPN, s\satCTLStar\stateFormula_1 \ \text{and}\ \runPN,s\satCTLStar\stateFormula_2 \\[1mm]
\runPN, s \satCTLStar \ctlExists\,\pathFormula					&\ \text{iff}\ &    \text{there \emph{exists} some flow chain suffix } \flowChain=t_0,p_0,\ldots \text{ of }\runPN\\[1mm]
																					&&\text{with } p_0\in s\text{ such that }\runPN,\trace(\flowChain)\satCTLStar\pathFormula\text{ holds for } s\not\subseteq\pl\\[1mm]
																					&&\text{and } \runPN,\trace_\stutterSymbol(s)\satCTLStar\pathFormula\text{ holds for } s\subseteq\pl\\[1mm]
\hline\\[-3mm]
\runPN,\trace \satCTLStar \stateFormula							&\ \text{iff}\ &	\runPN, \sigma(0) \satCTLStar \stateFormula\\[1mm]
\runPN,\trace \satCTLStar \lnot\, \pathFormula      				&\ \text{iff}\ & 	\text{not}\ \runPN, \trace \satCTLStar \pathFormula \\[1mm]
\runPN,\trace \satCTLStar \pathFormula_1 \land \pathFormula_2    &\ \text{iff}\ &    \runPN,\trace\satCTLStar\pathFormula_1\ \text{and}\ \runPN,\trace\satCTLStar\pathFormula_2\\[1mm]
\runPN,\trace \satCTLStar \ctlNext\,\pathFormula					&\ \text{iff}\ &    \runPN,\trace^1 \satCTLStar \pathFormula \\[1mm]
\runPN,\trace\satCTLStar\pathFormula_1\ctlUntil\, \pathFormula_2 &\ \text{iff}\ &    \text{there exists some } j\ge 0 \text{ with } \runPN,\trace^j \satCTLStar \pathFormula_2\text{ and} \\[1mm]
																					&&\text{for all }  0 \le i < j \text{ the following holds: } \runPN,\sigma^i \satCTLStar \pathFormula_1
\end{array}
\]
with atomic propositions \(a\in\mathit{AP}\), state formulas \(\stateFormula, \stateFormula_1\), and \(\stateFormula_2\), and path formulas \(\pathFormula,\pathFormula_1\), and \(\pathFormula_2\).
Note that since the formulas are evaluated on the runs of~\(\pNet\),
the branching is in the transitions and not in the places of \(\pNet\).

\subsection{Flow-\ctlStarText{}}
\label{sec:flowCTLStar}
Like in \cite{DBLP:conf/atva/FinkbeinerGHO19}, we use Petri nets with transits to enable reasoning 
about two separate timelines.
Properties defined on the run of the system concern the \emph{global} timeline and
allow to reason about the global behavior of the system like its general control or fairness.
Additionally, we can express requirements about the individual data flow like the access possibilities of people in buildings.
These requirements concern the \emph{local} timeline of the specific data flow.
In Flow-\ctlStarText{}, we can reason about these two parts
with \ltlText{} in the \emph{run} and with \ctlStarText{} in the \emph{flow} part of the formula. This is 
reflected in the following \emph{syntax}:
\[
 \phiRun  ::= \phiLTL \mid \phiRun_1 \land \phiRun_2
                  \mid \phiRun_1 \lor \phiRun_2 
                  \mid \phiLTL \rightarrow \phiRun
				  \mid \A\,\ctlStarFormula
\]
where \(\phiRun\), \(\phiRun_1\), \(\phiRun_2\) are Flow-\ctlStarText{} formulas,
\(\phiLTL\) is an \ltlText{} formula,
and \(\ctlStarFormula\) is a \ctlStarText{} formula.
We call \(\flowFormula=\A\,\ctlStarFormula\) \emph{flow formulas}
and all other subformulas \emph{run formulas}.

The \emph{semantics} of a Petri net with transits \(\petriNetFl\) satisfying a Flow-\ctlStarText{} formula \(\phiRun\)  
is defined over the covering firing sequences of its runs:
\[
\begin{array}{lll}
\pNet \models \phiRun		 								&\ \text{iff}\ & 		\text{for all runs } \runPN \text{ of }\pNet: \runPN\models \phiRun	  \\[1mm]
\runPN \models \phiRun									    &\ \text{iff}\ & 		\text{for all firing sequences } \firingSequence \text{ covering } \runPN: \ \runPN, \sigma(\firingSequence) \models \phiRun\\[1mm]
\runPN,\sigma\models \phiLTL 						&\ \text{iff}\ &    	\sigma\satLTL\phiLTL\\[1mm]
\runPN,\sigma\models \phiRun_1 \land \phiRun_2  		&\ \text{iff}\ &    	\runPN,\sigma\models\phiRun_1 \ \text{and}\ \runPN,\sigma\models\phiRun_2 \\[1mm]
\runPN,\sigma\models \phiRun_1 \lor \phiRun_2  		&\ \text{iff}\ &    	\runPN,\sigma\models\phiRun_1 \ \text{or}\ \runPN,\sigma\models\phiRun_2 \\[1mm]
\runPN,\sigma\models \phiLTL \rightarrow \phiRun     &\ \text{iff}\ & 		\runPN,\sigma\models\phiLTL \ \text{implies}\ \runPN,\sigma\models\phiRun\\[1mm]
\runPN,\sigma\models \A\,\ctlStarFormula				&\ \text{iff}\ &    	\text{for all flow chains }\flowChain\text{ of }\runPN:\runPN,\sigma(\flowChain) \satCTLStar\ctlStarFormula
\end{array}
\] 
Due to the covering of the firing sequences and the maximality constraint of the
flow chain suffixes, every behavior of the run is incorporated.
The operator~\(\A\) chooses flow chains rather than flow trees
as our definition is based on the common semantics of \ctlStarText{}
over paths.
Though it suffices to find one of the possibly infinitely many flow trees for each flow formula to
invalidate the subformula, checking the data flow while the control changes the system
complicates the direct expression of the model checking problem within a finite model.
In \refSection{mcFlowCTLStar}, we introduce a general reduction method
for a model with a finite state space.

\section{Example Specifications}
\label{sec:applications}
We illustrate Flow-\ctlStarText{} with examples from the literature on physical access control~\cite{DBLP:conf/csfw/TsankovDB16,DBLP:conf/est/GeepallaBD13}. 
Branching properties like permission and way-pointing are given as flow formulas, linear properties like fairness and maximality as run formulas. 

\subsection{Flow Formulas}
\tikzstyle{rectNode}=[rectangle,draw=black!75,fill=black!15,minimum size=4mm]
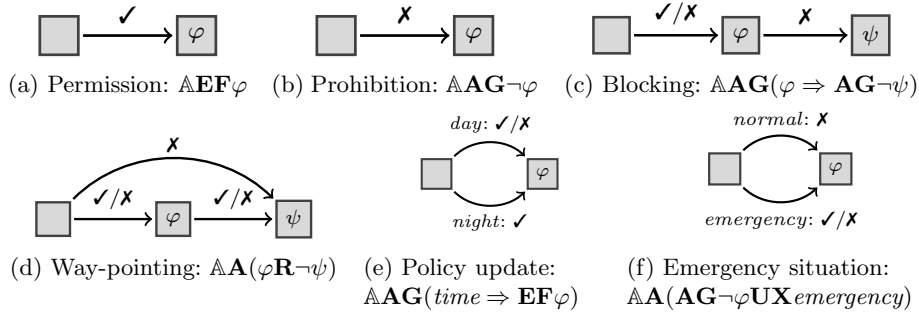
\begin{figure}[t]
\centering
	\begin{subfigure}[t]{0.29\textwidth}
		\centering
		\resizebox{0.75\textwidth}{!}{
		\begin{tikzpicture}
			\node[rectNode] (a) {};
			\node[rectNode, right=of a] (b) {\scriptsize\(\varphi\)};
			\draw[thick, shorten >=1pt, ->,shorten <=1pt] (a) -- node[above]{\scriptsize\cmark} (b);
		\end{tikzpicture}
		}
		\subcaption{Permission: $\A\ctlEF\varphi$}
		\label{fig:rulePermission}
	\end{subfigure}%
~
	\begin{subfigure}[t]{0.29\textwidth}
		\centering
		\resizebox{0.75\textwidth}{!}{
		\begin{tikzpicture}
			\node[rectNode] (a) {};
			\node[rectNode, right=of a] (b) {\scriptsize\(\varphi\)};
			\draw[thick, shorten >=1pt, ->,shorten <=1pt] (a) -- node[above]{\scriptsize\xmark} (b);
		\end{tikzpicture}
		}
		\subcaption{Prohibition: $\A\ctlAG\neg\varphi$}
		\label{fig:ruleProhibition}
	\end{subfigure}%
~
	\begin{subfigure}[t]{0.42\textwidth}
		\centering
		\resizebox{0.825\textwidth}{!}{
		\begin{tikzpicture}
			\node[rectNode] (a) {};
			\node[rectNode, right=of a] (b) {\scriptsize\(\varphi\)};
			\node[rectNode, right=of b] (c) {\scriptsize\(\psi\)};
			\draw[thick, shorten >=1pt, ->,shorten <=1pt] (a) -- node[above]{\scriptsize\cmark/\xmark} (b);
			\draw[thick, shorten >=1pt, ->,shorten <=1pt] (b) -- node[above]{\scriptsize\xmark} (c);
		\end{tikzpicture}
		}
		\subcaption{Blocking: $\A\ctlAG(\varphi\Rightarrow \ctlAG \neg \psi )$}
		\label{fig:ruleBlocking}
	\end{subfigure}%
\\
	\begin{subfigure}[t]{0.39\textwidth}
		\centering
		\resizebox{0.825\textwidth}{!}{
		\begin{tikzpicture}
			\node[rectNode] (a) {};
			\node[rectNode, right=of a] (b) {\scriptsize\(\varphi\)};
			\node[rectNode, right=of b] (c) {\scriptsize\(\psi\)};
			\draw[thick, shorten >=1pt, ->,shorten <=1pt] (a) -- node[above]{\scriptsize\cmark/\xmark} (b);
			\draw[thick, shorten >=1pt, ->,shorten <=1pt] (b) -- node[above]{\scriptsize\cmark/\xmark} (c);
			\draw[thick, shorten >=1pt, ->,shorten <=1pt] (a) edge[bend left=45] node[above,yshift=-0.5mm]{\scriptsize\xmark} (c);
		\end{tikzpicture}
		}
		\subcaption{Way-pointing: $\A\ctlAR{\varphi}{\neg\psi}$}
		\label{fig:ruleWaypointing}
	\end{subfigure}%
~
	\begin{subfigure}[t]{0.275\textwidth}
		\centering
		\resizebox{0.61\textwidth}{!}{
		\begin{tikzpicture}
			\node[rectNode] (a) {};
			\node[rectNode, right=of a] (b) {\scriptsize\(\varphi\)};
			\draw[thick, shorten >=1pt, ->,shorten <=1pt] (a) edge[bend left=45] node[above]{\scriptsize\emph{day}{:} \cmark/\xmark} (b);
			\draw[thick, shorten >=1pt, ->,shorten <=1pt] (a) edge[bend right=45] node[below]{\scriptsize\emph{night}{:} \cmark} (b);
		\end{tikzpicture}
		}
		\subcaption{Policy update:\\ $\A\ctlAG(\textit{time} \Rightarrow \ctlEF \varphi ) $}
		\label{fig:ruleTimeChange}
	\end{subfigure}%
~
	\begin{subfigure}[t]{0.335\textwidth}
		\centering
		\resizebox{0.61\textwidth}{!}{
		\begin{tikzpicture}
			\node[rectNode] (a) {};
			\node[rectNode, right=of a] (b) {\scriptsize\(\varphi\)};
			\draw[thick, shorten >=1pt, ->,shorten <=1pt] (a) edge[bend left=45] node[above]{\scriptsize\emph{normal}{:} \xmark} (b);
			\draw[thick, shorten >=1pt, ->,shorten <=1pt] (a) edge[bend right=45] node[below]{\scriptsize\emph{emergency}{:} \cmark/\xmark} (b);
		\end{tikzpicture}
		}
		\subcaption{Emergency situation:\\ $\A\ctlAU{\ctlAG\neg\varphi}{\ctlNext \textit{emergency}} $}
		\label{fig:ruleEmergency}
	\end{subfigure}%
	\caption{
		Illustrations for standard properties of physical access control are depicted. Gray boxes represent rooms and arrows represent directions of doors that can be opened~(\cmark), closed~(\xmark), or are not affected by the property (\cmark/\xmark).
	}
	\label{fig:rules}
\end{figure}
Figure~\ref{fig:rules} illustrates six typical specifications for physical access control~\cite{DBLP:conf/csfw/TsankovDB16,DBLP:conf/est/GeepallaBD13}.

\noindent
{\bf Permission.} 
Permission (cf.\ \refFig{rulePermission}) requires that a subformula $\varphi$ can be reached on one path ($\A \ctlEF \varphi $). 
In our running example, permission can be required for the \emph{hall} and the \emph{lab}. 
Permission can be extended as it requires reaching the subformula once. 
Persistent permission then requires that, on all paths, the subformula $\varphi$ can be repeatedly reached on a path ($\A\ctlAG\ctlEF \varphi $). 

\noindent
{\bf Prohibition.}
Prohibition (cf.\ \refFig{ruleProhibition}) requires that a subformula $\varphi$, for example representing a room, can never be reached on any path ($\A\ctlAG \neg \varphi $).
In our running example, closing the door to the \emph{kitchen} would satisfy prohibition for the \emph{kitchen}. 

\noindent
{\bf Blocking.}
Blocking (cf.\ \refFig{ruleBlocking}) requires for all paths globally that, after reaching subformula $\varphi$, the subformula $\psi$ cannot be reached ($\A\ctlAG (\varphi \Rightarrow \ctlAG \neg \psi)$).
This can be used to allow a new employee to only enter one of many labs.  

\noindent
{\bf Way-pointing.}
Way-pointing (cf.\ \refFig{ruleWaypointing}) ensures for all paths that subformula~$\psi$ can only be reached if $\varphi$ was reached before ($\A\ctlAR{\varphi}{\neg\psi}$). 
This can be used to enforce a mandatory security check when entering a building. 

\noindent
{\bf Policy update.}
A policy update (cf.\ \refFig{ruleTimeChange}) allows access to subformula $\varphi$ according to a time schedule ($\A\ctlAG(\textit{time} \Rightarrow \ctlEF \varphi ) $) with \emph{time} being a transition. 
This can be used to restrict access during the night.

\noindent
{\bf Emergency.}
An emergency situation (cf.\ \refFig{ruleEmergency}) can revoke the prohibition of subformula~$\varphi$ at an arbitrary time ($\A\ctlAU{\ctlAG\neg\varphi}{\ctlNext\textit{emergency}} $) with \emph{emergency} being a transition. 
An otherwise closed door could be opened to evacuate people. 
The next operator $\ctlNext$ is necessary because of the ingoing semantics of Flow-\ctlStarText{}.

\subsection{Run Formulas}

Flow formulas require behavior on the maximal flow of people in the building. 
Doors are assumed to allow passthrough in a fair manner. 
Both types of assumptions are expressed in Flow-\ctlStarText{} as run formulas. 

\noindent
{\bf Maximality.}
A run $\runPN$ is \emph{interleaving-maximal} if, whenever some transition is enabled, some transition will be taken:
 $\runPN \models \ltlAlways( \bigvee_{t \in \transitions} \pre{}{t} \rightarrow 
                        \bigvee_{t \in \transitions} \ltlNext t)$.
A run~$\runPN$ is \emph{concurrency-maximal} if, when a transition $t$ is from a moment on always
enabled, infinitely often a transition $t'$ (including $t$ itself) sharing a precondition with $t$ 
is taken:
 $\runPN \models \bigwedge_{t \in \transitions}
  ( \ltlEventually \ltlAlways\, \pre{}{t} \rightarrow 
    \ltlAlways \ltlEventually \bigvee_{{\small\begin{array}{c}
                             p\in \pre{}{t},
                             t'\in \post{}{p}
                           \end{array}}} t')$.

\noindent
{\bf Fairness.}
A run $\runPN$ is \emph{weakly fair} w.r.t.\ a transition $t$ if, whenever $t$ is always enabled after some point, $t$ is taken infinitely often:
{\(\runPN \models \ltlEventually \ltlAlways\, \pre{}{t} \rightarrow 
    \ltlAlways \ltlEventually t\)}.\\
A run $\runPN$ is \emph{strongly fair} w.r.t.\ $t$ if, whenever $t$ is enabled infinitely often, $t$ is taken infinitely often:
{\(\runPN \models \ltlAlways \ltlEventually\, \pre{}{t} \rightarrow \ltlAlways \ltlEventually t\)}.

\section{Model Checking Flow-\ctlStarText{} on Petri Nets with Transits}
\label{sec:mcFlowCTLStar}
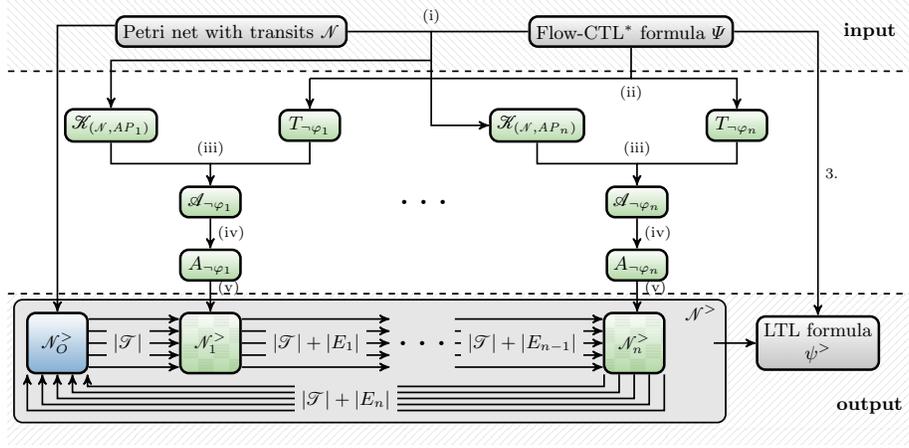
\begin{figure}[t]
\centering
 \scalebox{.8}{
 \begin{tikzpicture}[		
		->,
		thick,
		>=stealth',
		node distance=10mm and 20mm,
		lbl/.style={
		   align=center
		},
		elem/.style={
			rectangle,
			rounded corners,
			draw=black, 
			very thick,
		    text centered,
			minimum height=5mm,
		}
 ]
	\tikzstyle{myarrows}=[line width=1mm,-triangle 45,postaction={line width=2mm, shorten >=2mm, -}]

	\node[elem, bottom color=gray!50, top color=white,minimum height=5mm, minimum width=15mm, align=center, yshift=0mm]
					 (pnwt) {Petri net with transits \(\pNet\)};
	\node[elem, bottom color=gray!50, top color=white,minimum height=5mm, minimum width=8mm, align=center, right=of pnwt,xshift=10mm,yshift=0mm]
					 (phi) {Flow-\ctlStarText{} formula \(\phiRun\)};
	\node[right=of phi,xshift=-3mm,yshift=0mm] (input) {\textbf{input}};

	\node[elem, bottom color=cdc_Green!50, top color=white,minimum height=5mm, minimum width=10mm, align=center, below=of pnwt,xshift=-20mm]
					 (K1) {\(\kripke_{(\pNet,\AP_1)}\)};
	\node[elem, bottom color=cdc_Green!50, top color=white,minimum height=5mm, minimum width=10mm, align=center, right=of K1,yshift=0mm]
					 (T1) {\(\tree_{\lnot \ctlStarFormula_1}\)};
	\node[elem, bottom color=cdc_Green!50, top color=white,minimum height=5mm, minimum width=10mm, align=center, below=of K1,yshift=0mm]
			at ($(K1)!.5!(T1)$)	 (ABA1) {\(\aba_{\neg\ctlStarFormula_1}\)};
	\node[elem, bottom color=cdc_Green!50, top color=white,minimum height=5mm, minimum width=10mm, align=center, below=of ABA1,yshift=5mm]
		 			(AB1) {\(\ba_{\neg\ctlStarFormula_1}\)};

	\node[elem, bottom color=cdc_Green!50, top color=white,minimum height=5mm, minimum width=10mm, align=center, right=of K1,xshift=35mm]
					 (Kn) {\(\kripke_{(\pNet,\AP_n)}\)};
	\node[elem, bottom color=cdc_Green!50, top color=white,minimum height=5mm, minimum width=10mm, align=center, right=of Kn,yshift=0mm]
					 (Tn) {\(\tree_{\lnot \ctlStarFormula_n}\)};
	\node[elem, bottom color=cdc_Green!50, top color=white,minimum height=5mm, minimum width=10mm, align=center, below=of Kn,yshift=0mm]
			at ($(Kn)!.5!(Tn)$)	 (ABAn) {\(\aba_{\neg\ctlStarFormula_n}\)};
	\node[elem, bottom color=cdc_Green!50, top color=white,minimum height=5mm, minimum width=10mm, align=center, below=of ABAn,yshift=5mm]
		 			(ABn) {\(\ba_{\neg\ctlStarFormula_n}\)};

\node[rectangle, fill=white,minimum height=10mm,minimum width=6mm, align=center, xshift=0mm] at ($(ABA1)!.5!(ABAn)$) (dotsInner) {\huge\(\dots\)};

	\node[elem, below=of AB1, fill opacity=0.95,pattern=checkerboard, pattern color=cdc_Green!50!black, bottom color=cdc_Green!50, top color=white,minimum height=10mm,minimum width=10mm, align=center, yshift=5mm, xshift=0mm] (sub1) {\(\pNetMCSub{1}\)};
	\node[elem, left=of sub1, bottom color=cdc_Blue!50, top color=white,minimum height=10mm,,minimum width=10mm, align=center, yshift=0mm,xshift=5mm] (o) {\(\pNetMCO\)};
	\node[elem, right=of sub1, fill opacity=0.95,pattern=checkerboard, pattern color=cdc_Green!50!black, bottom color=cdc_Green!50, top color=white,minimum height=10mm,minimum width=10mm, align=center, xshift=40mm] (subn) {\(\pNetMCSub{n}\)};

\node[right=of subn,align=center,xshift=-18mm, yshift=5mm] (pnMC) {\(\pNetMC\)}; 

\node[elem, bottom color=gray!50, top color=white,minimum height=5mm, minimum width=8mm, align=center, right=of subn,xshift=-5mm]
					 (phiMC) {\ltlText{} formula\\\(\phiLTL\MC\)};
\node[below=of input,yshift=-47mm] (output) {\textbf{output}};

\node[rectangle, fill=white,minimum height=10mm,minimum width=6mm, align=center, xshift=0mm,fill=gray!20] at ($(sub1)!.5!(subn)$) (dots) {\huge\(\dots\)};

	\path (o) edge (sub1);
	\path ([yshift=2mm]o.east) edge ([yshift=2mm]sub1.west);
	\path ([yshift=4mm]o.east) edge ([yshift=4mm]sub1.west);
	\path ([yshift=-2mm]o.east) edge ([yshift=-2mm]sub1.west);
	\path ([yshift=-4mm]o.east) edge ([yshift=-4mm]sub1.west);

	\path (sub1) edge (dots);
	\path ([yshift=2mm]sub1.east) edge ([yshift=2mm]dots.west);
	\path ([yshift=4mm]sub1.east) edge ([yshift=4mm]dots.west);
	\path ([yshift=-2mm]sub1.east) edge ([yshift=-2mm]dots.west);
	\path ([yshift=-4mm]sub1.east) edge ([yshift=-4mm]dots.west);

	\path (dots) edge (subn);
	\path ([yshift=2mm]dots.east) edge ([yshift=2mm]subn.west);
	\path ([yshift=4mm]dots.east) edge ([yshift=4mm]subn.west);
	\path ([yshift=-2mm]dots.east) edge ([yshift=-2mm]subn.west);
	\path ([yshift=-4mm]dots.east) edge ([yshift=-4mm]subn.west);

	\node[rectangle, fill=white,minimum height=6mm,minimum width=6mm, align=center, xshift=-1mm,fill=gray!20] at ($(o)!.5!(sub1)$) (trO-1) {\(|\tr|\)};
	\node[rectangle, fill=white,minimum height=6mm,minimum width=6mm, align=center, xshift=0mm,fill=gray!20] at ($(sub1)!.5!(dots)$) (trO-2) {\(|\tr|+|E_{1}|\)};
	\node[rectangle, fill=white,minimum height=6mm,minimum width=6mm, align=center, xshift=-1mm,fill=gray!20] at ($(dots)!.5!(subn)$) (trO-3) {\(|\tr|+|E_{n-1}|\)};
	
	\draw[->] (subn.south) -- ++(0,-4mm) -- ([yshift=-4mm]o.south) -- (o.south);
	\draw[->] ([xshift=-2.5mm]subn.south) -- ++(0,-3mm) -- ([yshift=-3mm, xshift=2.5mm]o.south) -- ([xshift=2.5mm]o.south);
	\draw[->] ([xshift=-5mm]subn.south) -- ++(0,-2mm) -- ([yshift=-2mm, xshift=5mm]o.south) -- ([xshift=5mm]o.south);
	\draw[->] ([xshift=2.5mm]subn.south) -- ++(0,-5mm) -- ([yshift=-5mm, xshift=-2.5mm]o.south) -- ([xshift=-2.5mm]o.south);
	\draw[->] ([xshift=5mm]subn.south) -- ++(0,-6mm) -- ([yshift=-6mm, xshift=-5mm]o.south) -- ([xshift=-5mm]o.south);

	\node[rectangle, fill=white,minimum height=2mm,minimum width=2mm, align=center, yshift=-9.5mm, xshift=0mm,fill=gray!20] at ($(o)!.5!(subn)$) (trO-3) {\small\(|\tr|+|E_n|\)};

\node at ($(K1)!.5!(ABA1)$) (k1ABA1) {};
\node at ($(Kn)!.5!(ABAn)$) (knABAn) {};
\node at ($(pnwt)!.5!(T1)$) (pnwtT1) {};
\node at ($(pnwt)!.5!(phi)$) (pnwtphi) {};

\draw[->] ([yshift=1mm]pnwt.west) -- ([yshift=1mm]pnwt-|o) -- (o.north);

\draw[->] (phi.east) -- (phi-|phiMC) -- node[pos=0.5,right] {\scriptsize 3.} (phiMC.north);

\draw[->] (pnwt.east) -- node[pos=1,above] {\scriptsize(i)} ([xshift=0mm]pnwtphi.center) -- ([xshift=0mm]pnwtphi|-Kn) --
  (Kn);

\draw[-] (phi.west) -- ([xshift=0mm]pnwtphi.center);
\draw[->]  ([yshift=3mm]pnwtphi|-pnwtT1) -- ([yshift=3mm]pnwtT1-|K1) --
 ([yshift=0mm]K1);

\draw[->] (pnwtT1-|T1) -- (T1);

\draw[-] (phi.south) -- (phi|-pnwtT1)  -- node[pos=0,below] {\scriptsize(ii)} (pnwtT1-|T1);
\draw[->] (phi|-pnwtT1) -- (pnwtT1-|Tn) -- (Tn);

\draw[->] ([xshift=8mm]subn.east) -- (phiMC.west);

\draw[->] ([yshift=0mm]K1.south) -- (k1ABA1-|K1) --  (k1ABA1-|ABA1) -- ([yshift=0mm]ABA1.north);
\draw[-] ([yshift=0mm]T1.south) -- (k1ABA1-|T1) -- node[pos=1,above] {\scriptsize(iii)}  (k1ABA1-|ABA1);
\path ([yshift=0mm]ABA1.south) edge node[right] {\scriptsize(iv)}  ([yshift=0mm]AB1.north);
\path ([yshift=0mm]AB1.south) edge node[right,pos=0.2] {\scriptsize(v)}  ([yshift=0mm]sub1.north);

\draw[->] ([yshift=0mm]Kn.south) -- (knABAn-|Kn) --  (knABAn-|ABAn) -- ([yshift=0mm]ABAn.north);
\draw[-] ([yshift=0mm]Tn.south) -- (knABAn-|Tn) --node[pos=1,above] {\scriptsize(iii)}   (knABAn-|ABAn);
\path ([yshift=0mm]ABAn.south) edge node[right] {\scriptsize(iv)}  ([yshift=0mm]ABn.north);
\path ([yshift=0mm]ABn.south) edge node[right,pos=0.2] {\scriptsize(v)}  ([xshift=0.5mm]subn.north);

\draw[-, thick,dashed] ([xshift=-3mm,yshift=3mm]o.north west) -- ([xshift= 147mm,yshift=3mm]o.north west);
\draw[-, thick,dashed] ([xshift=-3mm,yshift=40mm]o.north west) -- ([xshift= 147mm,yshift=40mm]o.north west);

\begin{pgfonlayer}{background}
\fill [-, rectangle,fill=gray!15,pattern=north west lines,opacity=0.4,pattern color=gray]
([xshift=-3mm,yshift=40mm]o.north west)
-- ++(150mm,0)
-- ++(0,12mm)
-- ++(-150mm,0)
-- cycle;
\fill [-, rectangle,fill=gray!15,pattern=north east lines,opacity=0.4,pattern color=gray]
([xshift=-3mm,yshift=3mm]o.north west)
-- ++(150mm,0)
-- ++(0,-25mm)
-- ++(-150mm,0)
-- cycle;
\draw [-, rectangle,rounded corners,black,fill=gray!20]
 ([xshift=-2mm,yshift=2mm]o.north west) 
-- ([xshift=10mm,yshift=2mm]subn.north east) 
-- ([xshift=10mm,yshift=-8mm]subn.south east) 
-- ([xshift=-2mm,yshift=-8mm]o.south west) 
-- cycle;
\end{pgfonlayer}
\end{tikzpicture}
 }
	\caption{Overview of the model checking procedure: 
For a given safe Petri net with transits \(\pNet\) and a Flow-\ctlStarText{} formula \(\phiRun\), 
a standard Petri net \(\pNetMC\) and an \ltlText{} formula \(\phiLTL\MC\) are created:
For each flow subformula \(\A\,\ctlStarFormula_i\), 
create (i) a labeled Kripke structure \(\kripke_{(\pNet,\AP_i)}\)
and (ii) the alternating tree automaton \(\tree_{\lnot \ctlStarFormula_i}\), 
construct (iii) the alternating word automaton \(\aba_{\neg\ctlStarFormula_i}=\tree_{\neg \ctlStarFormula_i}\times\kripke_{(\pNet,\AP_i)}\),
and from that (iv) the B\"uchi automaton \(\ba_{\neg\ctlStarFormula_i}\) with edges \(E_i\),
which then (v) is transformed into a Petri net \(\pNetMC_i\).
These subnets are composed to a Petri net \(\pNetMC\) such that they get subsequently
triggered for every transition fired by the original net.
The constructed formula~\(\phiLTL\MC\) skips for the run part of \(\phiRun\) 
these subsequent steps and checks the acceptance of the guessed tree for each automaton.
The problem is then solved by checking \(\pNetMC\satLTL\phiLTL\MC\).
}
	\label{fig:overviewMC}
\end{figure}

We solve the model checking problem for a given Flow-\ctlStarText{} formula \(\phiRun\) and a safe Petri net with transits \(\pNet\)
in four steps:
\begin{enumerate}
\item For each flow subformula \(\A\,\ctlStarFormula_i\) of \(\phiRun\), 
a subnet \(\pNetMC_i\) is created via a sequence of automata constructions
which allows to guess a counterexample, i.e., a flow tree not satisfying \(\ctlStarFormula_i\),
and to check for its correctness.
\item The Petri net \(\pNet\MC\) is created by composing the subnets \(\pNetMC_i\)
to a copy of \(\pNet\) such that every firing of a transition subsequently triggers each subnet.
\item The formula \(\phiRun\MC\) is created such that
the subnets \(\pNetMC_i\) are adequately skipped for the run part of \(\phiRun\), and the 
flow parts are replaced by \ltlText{} formulas checking the acceptance of a run of the corresponding automaton.
\item \(\pNetMC\satLTL\phiRun\MC\) is checked to answer \(\pNet\models\phiRun\).
\end{enumerate}
The construction from a given
safe Petri net with transits \(\petriNetFl\)
and a Flow-\ctlStarText{} formula \(\phiRun\) with \(n\in\N\) flow subformulas \({\flowFormula}_i=\A\,\ctlStarFormula_i\)
with atomic propositions \(\AP_i\)
to a Petri net \(\pNetMC=(\plMC,\trMC,\flMC,\inhibitorFlMC,\initMC)\)
with inhibitor arcs (denoted by \(\inhibitorFlMC\))
and an \ltlText{} formula \(\phiRun\MC\)
is defined in the following sections.
More details and proofs can be found in Appendix~\ref{app:defsAndProofs}.
An \emph{inhibitor arc} connects a place \(p\) and a transition~\(t\) of a Petri net
such that \(t\) is only enabled when \(p\) is empty.
Figure~\ref{fig:overviewMC} gives a schematic overview of the procedure.

\subsection{Automaton Construction for Flow Formulas}
\label{sec:automatonConstruction}
In Step 1, we create for each flow subformula \(\A\,\ctlStarFormula_i\) of \(\phiRun\) with atomic propositions \(\AP_i\)
a nondeterministic B\"uchi automaton \(\ba_{\neg\ctlStarFormula_i}\)
which accepts a sequence of transitions of a given run if the corresponding flow tree
satisfies \(\neg\ctlStarFormula_i\).
This construction has four steps:
\begin{enumerate}
\item[(i)] Create the labeled Kripke structure \(\kripke_{(\pNet,\AP_i)}\) which, triggered by transitions \(t\in\tr\),
	 tracks	every flow chain of \(\pNet\). Each path corresponds to a flow chain.
\item[(ii)] Create the alternating tree automaton \(\tree_{\lnot \ctlStarFormula_i}\) 
	 for the negation of the \ctlStarText{} formula \(\ctlStarFormula_i\) and the set of directions
\(\D\subseteq\{0,\ldots,\mathtt{max}\{\size{\post{\tokenflow}{p,t}}-1 \with p\in\pl \wedge t\in\post{\pNet}{p}\}\}\)
	which accepts all \(2^{\AP_i}\)-labeled trees with nodes of degree in \(\D\) satisfying \(\lnot \ctlStarFormula_i\) \cite{DBLP:journals/jacm/KupfermanVW00}.
\item[(iii)] Create the alternating word automaton \(\aba_{\lnot \ctlStarFormula_i}=\tree_{\lnot \ctlStarFormula_i}\times\kripke_{(\pNet,\AP_i)}\) like in~\cite{DBLP:journals/jacm/KupfermanVW00}. 
\item[(iv)] Alternation elimination for \(\aba_{\lnot\ctlStarFormula_i}\) yields the nondeterministic B\"uchi automaton \(\ba_{\lnot \ctlStarFormula_i}\)~\cite{DBLP:journals/tcs/MiyanoH84,DBLP:conf/lpar/DaxK08}.
\end{enumerate}
Step~(ii) and Step~(iv) are well-established constructions.
For Step~(iii), we modify the construction of \cite{DBLP:journals/jacm/KupfermanVW00}
by applying the algorithm for the groups of equally labeled edges.
By this, we obtain an alternating word automaton with the alphabet \(A=\tr\cup\{\stutterSymbol\}\)
of the labeled Kripke structure rather than an alternating word automaton over a 1-letter alphabet.
This allows us to check whether the, by the input transition dynamically created, system
satisfies the \ctlStarText{} subformula~\(\ctlStarFormula_i\).

Step (i) of the construction creates the labeled Kripke structure~\(\kripke_{(\pNet,\AP_i)}=(\AP,S,S_0,L,A,R)\)
with a set of \emph{atomic propositions} \(\AP=\AP_i\),
a finite set of \emph{states} \(S=((\tr\cap\AP)\times\pl)\cup\pl\),
the \emph{initial states} \(S_0\subseteq S\),
the \emph{labeling function} \(L: S\to 2^{\AP}\),
the \emph{alphabet} \(A=\tr\cup\{\stutterSymbol\}\),
and the \emph{labeled transition relation} \(R\subseteq S\times A\times S\).
The Kripke structure serves (in combination with the tree automaton)
for checking the satisfaction of a flow tree of a given run.
Hence, the states track the current place of the considered chain of the tree
and additionally, when the transition extending the chain into the place occurs in the formula,
also this ingoing transition.
The initial states \(S_0\) are either the tuples of transitions \(t_j\) and places \(p_j\)
which start a flow chain, i.e., all \((t_j,p_j)\in\tr\times\pl\) with \((\startfl,p_j)\in\tokenflow(t_j)\)
when \(t_j\in\AP\) or only the place \(p_j\) otherwise.
The labeling function \(L\) labels the states with its components.
The transition relation \(R\) connects the states with respect to the transits,
connects each state \((t,p)\in\kstates\) with \(\stutterSymbol\)-labeled edges 
to the state \(p\in\kstates\),
and loops with \(\stutterSymbol\)-labeled edges in states \(s\in\pl\)
to allow for the stuttering of finite chains.

\begin{lemma}[Size of the Kripke Structure]
\label{lem:sizeKripkeStructure}
The constructed Kripke structure \(\kripke_{(\pNet,\AP_i)}\) has \(\oclass{\size{\AP_i\cap\tr}\cdot\size{\pNet}+\size{\pNet}}\) states 
and \(\oclass{\size{\pNet^3}}\) edges.
\end{lemma}
Note that the number of edges stems from the number of transits \((p,t,q)\in\pl\times\tr\times\pl\) 
used in the Petri net with transits \(\pNet\).

The size of the B\"uchi automaton is dominated by the tree automaton construction 
and the removal of the alternation. Each construction adds one exponent for \ctlStarText{}.
\begin{lemma}[Size of the B\"uchi Automaton]
\label{lem:sizeBuchiAutomaton}
The size of the B\"uchi automaton~\(\ba_{\lnot \ctlStarFormula_i}\)
is in \(\oclass{2^{2^{\size{\ctlStarFormula}}\cdot\size{\pNet}^3}}\)
for specifications \(\ctlStarFormula_i\) in \ctlStarText{} and in 
\(\oclass{2^{\size{\ctlStarFormula}\cdot\size{\pNet}^3}}\) for specifications in \ctlText{}.
\end{lemma}

\subsection{From Petri Nets with Transits to Petri Nets}
\label{sec:pnwt2pn}
In Step 2, we construct for the Petri net with transits \(\pNet\) and the B\"uchi automata~\(\ba_{\neg\ctlStarFormula_i}\)
for each flow subformula \({\flowFormula}_i=\A\,\ctlStarFormula_i\) of \(\phiRun\), 
a Petri net \(\pNetMC\) by composing a copy of \(\pNet\) (without transits), denoted by \(\pNetMCO\),
to subnets \(\pNetMCSub{i}\) corresponding to \(\ba_{\neg\ctlStarFormula_i}\)
such that each copy is sequentially triggered when a transition of \(\pNetMCO\) fires.
The subnet \(\pNetMCSub{i}\), when triggered by transitions \(t\in\tr\),
guesses nondeterministically the violating flow tree of the operator \(\A\)
and simulates \(\ba_{\neg\ctlStarFormula_i}\).
Thus, a token from the initially marked place \(\initsub_i\) is moved 
via a transition for each transition \(t\in\tr\) starting a flow chain
to the place corresponding to the initial state of \(\ba_{\neg\ctlStarFormula_i}\).
For each state \(s\) of \(\ba_{\neg\ctlStarFormula_i}\), we have a place~\(\subnet{s}_i\),
and, for each edge \((s,l,s')\), a transition labeled by \(l\)
which moves the token from \(\subnet{s}_i\) to \(\subnet{s'}_i\).

There are two kinds of stutterings: \emph{global} stuttering for finite
runs and \emph{local} stuttering for finite flow chains.
To guess the starting time of both stutterings,
there is an initially marked place \(\normalMode\), a place \(\stutteringMode\),
and a transition which can switch from \emph{normal} to \emph{stuttering} mode
for the global stuttering in~\(\pNetMCO\) and for the local stutterings in each subnet \(\pNetMC_i\)
(denoted by \(\subnet{\normalMode}\),\(\subnet{\stutteringMode}\)).
The original transitions of~\(\pNetMCO\) and the transitions of a subnet \(\pNetMC_i\) corresponding to 
a transition \(t\in\tr\) depend on the normal mode.
The \(\stutterSymbol\)-labeled transitions (used for global stuttering)
of the subnet depend on the stuttering mode.
To enable local stuttering, we add, for each edge \(e=(s,\stutterSymbol,s')\) of \(\ba_{\neg\ctlStarFormula_i}\),
a transition \(t\MC\) for each transition \(t\in\tr\) for which no edge \((s,t,s'')\) exists in \(\ba_{\neg\ctlStarFormula_i}\).
These transitions depend on the stuttering mode and move the token according to 
their corresponding edge~\(e\).

The original part \(\pNetMCO\) and the subnets \(\pNetMCSub{i}\) are 
connected in a sequential manner.
The net \(\pNetMCO\) has an initially marked activation place \(\act{o}\) in the preset of each transition,
the subnets have one activation place \(\subnet{\act{t}}\) in the preset of every
transition \(t\MC\) corresponding to a transition \(t\in\tr\) (normal as well as stuttering).
The transitions move the activation token to the corresponding places of the next subnet (or back to \(\pNetMCO\)).
To ensure the continuation even though the triggering transition does not extend the current flow tree
(e.g., because it is a concurrent transition of the run),
there is a skipping transition for each transition \(t\in\tr\) which moves the activation token
when none of the states having a successor edge labeled with \(t\) are active.
For the global stuttering, each subnet has an activation place \(\subnet{\act{\stutterSymbol}}_i\),
in which an additional transition \(t_\stutterSymbol\) in \(\pNetMCO\) puts the active token 
if the stuttering mode of \(\pNetMCO\) is active.
Each \(\stutterSymbol\)-labeled transition of the subnets moves this token to the next subnet (or back to \(\pNetMCO\)).

By that, we can check the acceptance of each \(\ba_{\neg\ctlStarFormula_i}\)
by checking if the subnet infinitely often reaches any places corresponding 
to a B\"uchi state of \(\ba_{\neg\ctlStarFormula_i}\).
This and only allowing to correctly guess the time point of the stutterings
is achieved with the formula described in \refSection{flowCTLStar2LTL}.
A formal definition is given in \refDef{pnwt2pn} in App.~\ref{app:defsAndProofs}.
The size of the constructed Petri net is dominated by the respective single- or double-exponential size
of the nondeterministic B\"uchi automata.
\begin{lemma}[Size of the Constructed Net]
\label{lem:sizeMCNet}
The constructed Petri net with inhibitor arcs \(\pNetMC\) for a Petri net with transits \(\pNet\) and
\(n\) nondeterministic B\"uchi automata
\(\ba_{\lnot \ctlStarFormula_i}=(\tr\cup\{\stutterSymbol\}, Q_i, I_i, E_i,F_i)\)
has \(\oclass{\size{\pNet}\cdot n + \size{\pNet} +\sum_{i=1}^n\size{Q_i}}\)
places and \(\oclass{\size{\pNet}^2\cdot n + \size{\pNet}+\sum_{i=1}^n\size{E_i}+\size{\pNet}\cdot\sum_{i=1}^n\size{Q_i}}\)
transitions.
\end{lemma}

\subsection{From Flow-\ctlStarText{} Formulas to \ltlText{} Formulas}
\label{sec:flowCTLStar2LTL}
The formula transformation from a given Flow-\ctlStarText{} formula \(\phiRun\)
and a Petri net with transits \(\pNet\) into an \ltlText{} formula (Step 3)
consists of three parts:

First, we substitute the flow formulas \({\flowFormula}_i=\A\,\ctlStarFormula_i\) 
with the acceptance check of the corresponding automaton \(\ba_{\neg\ctlStarFormula_i}\), i.e.,
we substitute \({\flowFormula}_i\) with \(\lnot\ltlAlways\ltlEventually\bigvee_{b\in F_i} \subnet{b}_i\)
for the B\"uchi states \(F_i\) of \(\ba_{\neg\ctlStarFormula_i}\).

Second, the sequential manner of the constructed net \(\pNetMC\) requires an adaptation of the
run part of \(\phiRun\).
For a subformula \(\phiLTL_1\ltlUntil\phiLTL_2\) with transitions \(t\in\tr\) as atomic propositions
or a subformula \(\ltlNext\,\phiLTL\) in the run part of \(\phiRun\),
the sequential steps of the subnets have to be skipped.
Let \(\trMC_O\) be the transition of the original copy~\(\pNetMC_O\),
\(\trMC_i\) the transitions of the subnet \(\pNetMC_i\),
\(\tr_{\skipT_i}\) the transitions of the subnet \(\pNetMC_i\) 
which skip the triggering of the automaton in the normal mode, and
\(t_{\normalMode\to\stutteringMode}\) the transition switching \(\pNetMC_O\) from normal to stuttering mode.
Then, because of the ingoing semantics, we can can select all states corresponding
to the run part with \(\mathtt{M}=\bigvee_{t\in\trMC_O\setminus\{t_{\normalMode\to\stutteringMode}\}} t\) together with the 
initial state \(\mathtt{i}=\neg\bigvee_{t\in\trMC} t\).
Hence, we replace each subformula \(\phiLTL_1\ltlUntil\phiLTL_2\)
containing transitions \(t\in\tr\) as atomic propositions with
\(((\mathtt{M}\vee\mathtt{i})\rightarrow\phiLTL_1)\ltlUntil((\mathtt{M}\vee\mathtt{i})\rightarrow\phiLTL_2)\)
from the inner- to the outermost occurrence.  
For the next operator, the second state is already the correct next state of the initial state
also in the sense of the global timeline of \(\phiLTL\MC\).
For all other states belonging to the run part (selected by the until construction above), 
we have to get the next state and then skip all transitions of the subnet.
Thus, we replace each subformula \(\ltlNext\, \phiLTL\) with
\(\mathtt{i}\rightarrow\ltlNext\,\phiLTL
\wedge
\neg\mathtt{i}\rightarrow\ltlNext(\bigvee_{t\in\trMC\setminus\trMC_O}t\ltlUntil\bigvee_{t'\in\trMC_O}t'\wedge\phiLTL)\)
from the inner- to the outermost occurrence.

Third, we have to ensure the correct switching into the stuttering mode.
By \(\mathtt{skip_i}=\neg\ltlEventually\ltlAlways((\bigvee_{t\in\trMC_i} t)\rightarrow(\bigvee_{t'\in\tr_{\skipT_i}} t'))\)
a subnet is enforced to switch into its stuttering mode if necessary.
If it wrongly selects the time point of the global stuttering, the run stops. 
Hence, we obtain the formula 
\(\phiLTL\MC=(\left(\ltlAlways\ltlEventually\act{o}\right)\wedge\bigwedge_{i\in\{1,\ldots,n\}}\mathtt{skip_i})\rightarrow\phiLTL\)
by only selecting the runs where the original part is infinitely often activated
and each subnet chooses its stuttering mode correctly.

Since the size of the formula depends on the size of the constructed Petri net~\(\pNetMC\),
it is also dominated by the B\"uchi automaton construction.
\begin{lemma}[Size of the Constructed Formula]
\label{lem:sizeFormula}
The size of the constructed formula \(\phiLTL\MC\) is double-exponential for specifications given in \ctlStarText{} and single-exponential
for specifications in \ctlText{}.
\end{lemma}
We can show that the construction of the net and the formula adequately fit together
such that the additional sequential steps of the subnets are skipped in the formula
and the triggering of the subnets simulating the B\"uchi automata as well as the stuttering
is handled properly.
\begin{lemma}[Correctness of the Transformation]
\label{lem:correctnessTransformation}
For a Petri net with transits \(\pNet\)
and a Flow-\ctlStarText{} formula \(\phiRun\),
there exists a safe Petri net \(\pNetMC\) with inhibitor arcs
and an LTL formula \(\phiRun\MC\)
such that \(\pNet\models\phiRun\) iff \(\pNetMC\satLTL\phiRun\MC\).
\end{lemma}

The complexity of the model checking problem of Flow-\ctlStarText{} is dominated by the 
automata constructions for the \ctlStarText{} subformulas.
The need of the alternation removal (Step (iv) of the construction)
is due to the checking of branching properties 
on structures chosen by linear properties.
In contrast to standard \ctlStarText{} model checking on a static Kripke structure,
we check on Kripke structures dynamically created for specific runs.
\begin{theorem}
\label{theo:complexityMC}
A safe Petri net with transits \(\pNet\) can be checked against a Flow-\ctlStarText{} formula \(\phiRun\) in 
triple-exponential time in the size of \(\pNet\) and \(\phiRun\).
For a Flow-\ctlText{} formula \(\phiRun'\), the model checking algorithm runs in double-exponential time in 
the size of \(\pNet\) and \(\phiRun'\).
\end{theorem}
Note that a single-exponential time algorithm for Flow-LTL is presented in \cite{DBLP:conf/atva/FinkbeinerGHO19}.

\section{Related Work}
\label{sec:related-work}
There is a large body of work on physical access control: 
Closest to our work are \emph{access nets}~\cite{DBLP:conf/vmcai/FrohardtCS11} which extend Petri nets with mandatory transitions to make people leave a room at a policy update. 
Branching properties can be model checked for a fixed number of people in the building. 
Fixing the number of people enables explicit interaction between people. 
In logic-based access-control frameworks, credentials are collected from distributed components to open policy enforcement points according to the current policy~\cite{DBLP:conf/sp/BauerGR05,DBLP:conf/esorics/BauerGR07}.
Techniques from networking can be applied to physical access control to detect redundancy, shadowing, and spuriousness in policies~\cite{DBLP:conf/crisis/FitzgeraldTFO12}. 
Our model prevents such situations by definition as a door can be either open or closed for people with the same access rights. 

A user study has been carried out to identify the limitations of physical access control for real-life professionals~\cite{DBLP:conf/chi/BauerCRRV09}. 
Here, it was identified that policies are made by multiple people which is a problem our approach of global control solves.
Types of access patterns are also studied~\cite{DBLP:conf/dbsec/FernandezBDL07,DBLP:conf/est/GeepallaBD13,DBLP:conf/csfw/TsankovDB16}: Access policies according to time schedules and emergencies, access policies for people without RFID cards, and dependent access are of great importance. 
The first and the third problem are solvable by our approach and the second one seems like an intrinsic problem to physical access control. 
Policies for physical access control can be synthesized if no policy updates are necessary~\cite{DBLP:conf/csfw/TsankovDB16}. 
It is an interesting open question whether policy updates can be included in the synthesis of access policies.

\section{Conclusion}
\label{sec:conclusion}
We present the first model checking approach for the verification of physical access control with policy updates under fairness assumptions and with an unbounded number of people. 
Our approach builds on Petri nets with transits which superimpose a transit relation onto the flow relation of Petri nets
to differentiate between data flow and control.
We introduce Flow-\ctlStarText{} to specify branching properties on the data flow and linear properties on the control in Petri nets with transits. 
We outline how Petri nets with transits can model physical access control with policy updates and how Flow-\ctlStarText{} can specify properties on the behavior before, during, and after updates including fairness and maximality.
To solve the model checking problem, we reduce the model checking problem of Petri nets with transits against Flow-\ctlStarText{} via automata constructions to the model checking problem of Petri nets against LTL.
In the future, we plan to evaluate our approach in a tool implementation and a corresponding case study. 
We can build on our tool \textsc{AdamMC}~\cite{DBLP:conf/cav/FinkbeinerGHO20} for Petri nets with transits and Flow-LTL.

\bibliographystyle{splncs04}
\bibliography{ms}

\begin{thebibliography}{10}
\providecommand{\url}[1]{\texttt{#1}}
\providecommand{\urlprefix}{URL }
\providecommand{\doi}[1]{https://doi.org/#1}

\bibitem{DBLP:conf/chi/BauerCRRV09}
Bauer, L., Cranor, L.F., Reeder, R.W., Reiter, M.K., Vaniea, K.: Real life
  challenges in access-control management. In: Proc.\ of {CHI} (2009)

\bibitem{DBLP:conf/sp/BauerGR05}
Bauer, L., Garriss, S., Reiter, M.K.: Distributed proving in access-control
  systems. In: Proc.\ of S{\&}P (2005)

\bibitem{DBLP:conf/esorics/BauerGR07}
Bauer, L., Garriss, S., Reiter, M.K.: Efficient proving for practical
  distributed access-control systems. In: Proc.\ of {ESORICS} (2007)

\bibitem{DBLP:reference/mc/2018}
Clarke, E.M., Henzinger, T.A., Veith, H., Bloem, R. (eds.): Handbook of Model
  Checking. Springer (2018)

\bibitem{DBLP:conf/lpar/DaxK08}
Dax, C., Klaedtke, F.: Alternation elimination by complementation (extended
  abstract). In: Proc. of {LPAR} (2008)

\bibitem{DBLP:journals/acta/Engelfriet91}
Engelfriet, J.: Branching processes of {P}etri nets. Acta Inf.  \textbf{28}(6)
  (1991)

\bibitem{DBLP:series/eatcs/EsparzaH08}
Esparza, J., Heljanko, K.: Unfoldings -- {A} Partial-Order Approach to Model
  Checking. Springer (2008)

\bibitem{DBLP:conf/dbsec/FernandezBDL07}
Fern{\'{a}}ndez, E.B., Ballesteros, J., Desouza{-}Doucet, A.C.,
  Larrondo{-}Petrie, M.M.: Security patterns for physical access control
  systems. In: Proc.\ of Data and Applications Security XXI (2007)

\bibitem{DBLP:conf/atva/FinkbeinerGHO19}
Finkbeiner, B., Gieseking, M., Hecking{-}Harbusch, J., Olderog, E.: Model
  checking data flows in concurrent network updates. In: Proc.\ of {ATVA}
  (2019)

\bibitem{DBLP:conf/cav/FinkbeinerGHO20}
Finkbeiner, B., Gieseking, M., Hecking{-}Harbusch, J., Olderog, E.: {AdamMC}: A
  model checker for {P}etri nets with transits against {Flow-LTL}. In: Proc.\
  of {CAV} (2020)

\bibitem{atva20}
Finkbeiner, B., Gieseking, M., Hecking{-}Harbusch, J., Olderog, E.: Model
  checking branching properties on {P}etri nets with transits. In: Proc. of
  {ATVA} (2020)

\bibitem{DBLP:conf/crisis/FitzgeraldTFO12}
Fitzgerald, W.M., Turkmen, F., Foley, S.N., O'Sullivan, B.: Anomaly analysis
  for physical access control security configuration. In: Proc.\ of {CRiSIS}
  (2012)

\bibitem{DBLP:conf/vmcai/FrohardtCS11}
Frohardt, R., Chang, B.E., Sankaranarayanan, S.: Access nets: Modeling access
  to physical spaces. In: Proc.\ of {VMCAI} (2011)

\bibitem{DBLP:conf/est/GeepallaBD13}
Geepalla, E., Bordbar, B., Du, X.: Spatio-temporal role based access control
  for physical access control systems. In: Proc.\ of {EST} (2013)

\bibitem{jensen92}
Jensen, K.: Coloured Petri Nets: Basic Concepts, Analysis Methods and Practical
  Use, Volume 1. Springer (1992)

\bibitem{DBLP:journals/jacm/KupfermanVW00}
Kupferman, O., Vardi, M.Y., Wolper, P.: An automata-theoretic approach to
  branching-time model checking. J. {ACM}  \textbf{47}(2) (2000)

\bibitem{DBLP:journals/tcs/MiyanoH84}
Miyano, S., Hayashi, T.: Alternating finite automata on omega-words. Theor.
  Comput. Sci.  \textbf{32} (1984)

\bibitem{DBLP:books/sp/Reisig85a}
Reisig, W.: Petri Nets: An Introduction. Springer (1985)

\bibitem{DBLP:conf/csfw/TsankovDB16}
Tsankov, P., Dashti, M.T., Basin, D.A.: Access control synthesis for physical
  spaces. In: Proc.\ of {CSF} (2016)

\bibitem{DBLP:journals/internet/WelbourneBCGRRBB09}
Welbourne, E., Battle, L., Cole, G., Gould, K., Rector, K., Raymer, S.,
  Balazinska, M., Borriello, G.: Building the internet of things using {RFID:}
  the {RFID} ecosystem experience. {IEEE} Internet Comput.  \textbf{13}(3)
  (2009)

\end{thebibliography}

\appendix
\section*{Appendix}
\section{Definitions for Petri Nets and Unfoldings}
\label{app:pn}
In this section of the appendix we recall some definitions for safe Petri nets and unfoldings.

Let \(\pNet=(\pl,\tr,\fl,\init)\) be a safe Petri net.
Thus, every \emph{marking} \(M\) of \(\pNet\) is a set (rather than a multiset) of places, i.e., \(M\subseteq\pl\).
We define the \textit{preset} of a node~$x$ from $\pNet$
as $\pre{\pNet}{x} = \{y \in \pl \cup \tr \mid (y,x)\in\fl\}$ 
and the postset as $\post{\pNet}{x} = \{y \in \pl \cup \tr \mid (x,y)\in\fl\}$. 
A transition \(t\in\tr\) is \emph{enabled} at a marking \(M\) iff \(\pre{\pNet}{t}\subseteq M\).
\emph{Firing} an enabled transition \(t\in\tr\) at a marking~\(M\)
yields the successor marking~\(M'=(M\setminus\pre{\pNet}{t})\cup\post{\pNet}{t}\). 
We denote this \emph{firing relation} by \(M\firable{t} M'\).
The \emph{interleaving semantics} of a safe Petri net 
has the \emph{reachable markings}
\(\reach(\pNet)=\init\cup\{M_n\subseteq\pl\with\exists t_1,\ldots,t_n\in\tr:\init\firable{t_1}M_1\firable{t_2}\cdots
\firable{t_n}M_n\}\)
as states and connects them according to the firing relation.

The following paragraphs introduce the \emph{unfolding} of a safe Petri net,
a \emph{true concurrency semantics} obtained by unfolding the behavior of the net into a tree.
The nodes of~\(\pNet\) can be partially ordered
by their causal dependencies.
For two nodes \(x,y\in\pl\cup\tr\) we call \(x\) a \emph{causal predecessor} of \(y\),
written \(x<y\),
iff \(x\,\fl^+\,y\) holds, i.e., 
\(y\) can be reached by following directed arcs from~\(x\).
We write \(x\leq y\) iff \(x<y\) or \(x=y\).
The nodes \(x\) and \(y\) are \emph{causally related} iff \(x\leq y\) or \(y\leq x\) holds. 
They are in \emph{conflict}, written \(\conflict{x}{y}\),
iff there is a place \(p\in\pl\setminus\{x,y\}\) and two transitions \(t_1,t_2\in \post{\pNet}{p}\)
with \(t_1\neq t_2\) such that \(t_1\leq x\) and \(t_2\leq y\) holds.
If they are neither in conflict nor causally related,
we call the nodes \emph{concurrent}.
A set of places \(X\subseteq\pl\) is called \emph{concurrent} iff
all places are pairwise concurrent.

An \emph{occurrence net} is a Petri net \(\pNet=(\pl,\tr,\fl,\init)\)
which represents the occurrences of transitions with their conflicts 
and causal dependencies with the following constraints:
(i)~\(\forall p\in\pl: |\pre{\pNet}{p}|\leq 1\),
(ii)~\(\forall t\in\tr: \neg(\conflict{t}{t})\),
(iii)~\(\forall x\in\pl\,\cup\,\tr: \neg(x<x)\),
(iv)~\(\forall x\in\pl\,\cup\,\tr: |\{y\in\pl\cup\tr\with y < x\}| < \infty\), and
(v)~\(\init=\{p\in\pl\with\pre{\pNet}{p}=\emptyset\}\).
Thus, each place has only one ingoing transition,
no transition is in self-conflict,
the flow relation is acyclic,
the relation \(<\) is well-founded, i.e., does not contain any infinitely decreasing sequence,
and the initial marking consists of exactly the places not having any predecessor.
We call an occurrence net a \emph{causal net}, when further each place has at most 
one transition as successor, i.e., (vi)~\(\forall p\in\pl:|\post{\pNet}{p}|\leq 1\) holds.

Let \(\pNet_1=(\pl_1,\tr_1,\fl_1,\init_1)\) and \(\pNet_2=(\pl_2,\tr_2,\fl_2,\init_2)\)
be two Petri nets.
We call \(\pNet_1\) a \emph{subnet} of \(\pNet_2\) iff \(\pl_1\subseteq\pl_2\), \(\tr_1\subseteq\tr_2\),
\(\fl_1\subseteq\fl_2\), and \(\init_1=\init_2\) holds.
A \emph{homomorphism} from \(\pNet_1\) to \(\pNet_2\) is a mapping \(h:\pl_1\cup\tr_1\to\pl_2\cup\tr_2\)
satisfying the following constraints:
(i)~\(h(\pl_1)\subseteq \pl_2\) and \(h(\tr_1)\subseteq\tr_2\) and 
(ii)~\(\forall t\in\tr_1: h(\pre{\pNet_1}{t})=\pre{\pNet_2}{h(t)}\wedge h(\post{\pNet_1}{t})=\post{\pNet_2}{h(t)}\),
with the component-wise application of the homomorphism to a set \(X\subseteq\pl_1\cup\tr_1\), i.e,
\(h(X)=\{h(x)\with x\in X\}\).
This means as homomorphism preserves the types of the nodes and the pre- and postconditions of the transitions.
We call \(h\) \emph{initial} iff also (iii)~\(h(\init_1)=\init_2\).
We assume the elements of a superscripted Petri net \(\pNet^X\)
implicitly to be superscripted accordingly, i.e., \(\pNet^X=(\pl^X,\tr^X,\fl^X,\init^X)\).

A \emph{branching process} \(\beta=(\pNet^U,\lambda^U)\) of a Petri net \(\pNet\) consists
of an occurrence net \(\pNet^U\) and a homomorphism \(\lambda^U:\pl^U\cup\tr^U\to\pl\cup\tr\) such that
\(\forall t_1,t_2 \in \tr^U : (\pre{\pNet}{t_1} = \pre{\pNet}{t_2} \wedge \lambda^U(t_1) = \lambda^U(t_2)) \Rightarrow t_1 = t_2\)
holds. This means \(\lambda^U\) is injective on transitions with the same preset.
We call \(\beta\) \emph{initial} iff \(\lambda^U\) is initial.
A branching process \(\beta_R=(\pNet^R,\rho)\) of \(\pNet\)
is called \emph{(concurrent) run} of \(\pNet\) iff \(\pNet^R\) is a causal net
and called an \emph{initial (concurrent) run} iff furthermore \(\rho\) is an initial homomorphism.
A run formalizes a single concurrent execution of the net.
For a function \(h\) let \(h\mid_X\) restrict the domain of \(h\) to the set \(X\).
A branching process \(\beta_1=(\pNet_1,\lambda_1)\) is called a \emph{subprocess} of
a branching process \(\beta_2=(\pNet_2,\lambda_2)\) iff \(\pNet_1\) is a subnet of \(\pNet_2\)
and \(\lambda_1={\lambda_2}\mid_{\pl_1\cup\tr_1}\).

An \emph{unfolding} of a net \(\pNet\)
is an initial branching process \(\beta=(\pNet^U,\lambda^U)\) of~\(\pNet\)
which has a transition \(t^U\) labeled with \(t\) whenever there is a transition \(t\in\tr\)
which can extend the unfolding:
\(\forall t\in\tr, C\subseteq\pl^U: C \text{ concurrent} \wedge \lambda^U(C)=\pre{\pNet}{t}\Rightarrow\exists t^U\in\tr^U:\pre{\pNet^U}{t^U}=C\wedge \lambda^U(t^U)=t\).
An unfolding is unique up to isomorphism.
Each run of \(\pNet\) is a subprocess of an unfolding \(\beta\).

We lift the transit relation of a Petri net with transits to any branching process \(\beta=(\pNet^U,\lambda^U)\)
and thereby obtain notions of \emph{runs} and \emph{unfoldings} for \emph{Petri nets with transits}.
The \emph{transit relation} \(\tfl^U\) of a branching process with transits \(\beta\)
of a Petri net with transits \(\pNet\) is defined as follows:
For any \(t\in\tr^U\), we define  
\(\tfl^U(t) \subseteq (\pre{\pNet^U}{t} \cup \{\startfl\}) \times \post{\pNet^U}{t}\)
such that \((p,q)\in\tfl^U(t)\Leftrightarrow(\lambda^U(p),\lambda^U(q))\in\tfl(\lambda^U(t))\) holds for all \(p,q\in\pl^U\).

\section{Formal Definitions and Proofs of the Model Checking Procedure}
\label{app:defsAndProofs}
In this section of the appendix we provide the formal definitions and proofs for \refSection{mcFlowCTLStar}.
We fix a Petri net with transits \(\petriNetFl\) and a Flow-\ctlStarText{} formula~\(\phiRun\)
with \(n\in\N\) flow subformulas \({\flowFormula}_i=\A\,\ctlStarFormula_i\) with atomic
propositions \(\AP_i\) throughout the section.

We formally define the labeled Kripke structure of which the unwinding triggered by
a firing sequence corresponds to one flow tree of a run of \(\pNet\).

\begin{definition}[Kripke Structure]
\label{def:kripkeStructure}
We construct the labeled Kripke structure 
\(\kripke_{(\pNet,\AP_i)}=(\AP,\katoms, \kstates,\kinit,\klab,\krelLab)\)
with
\begin{itemize}
\item the finite set of \emph{atomic propositions} \(\AP=\AP_i\),
\item the finite set of \emph{states} \(\kstates\subseteq((\tr\cap\AP)\times\pl)\cup\pl\),
\item the finite set of \emph{initial states} \(\kinit\subseteq \kstates\),
\item the \emph{labeling function} \(\klab: \kstates\to 2^{\AP}\),
\item the \emph{alphabet} \(\katoms=\tr\cup\{\stutterSymbol\}\),
\item the \emph{labeled transition relation} \(\krelLab\subseteq \kstates\times \katoms\times \kstates\).
\end{itemize}
The \emph{initial states} \(\kinit=\{(t,p)\in\tr\times\pl\with \exists t\in\tr\cap\AP: (\startfl,p)\in\tokenflow(t)\}\cup\{p\in\pl\with \exists t\in\tr\setminus\AP: (\startfl,p)\in\tokenflow(t)\}\)
correspond to all tuples of transitions \(t\in\AP\) and places (or only to the places) which start a data flow in \(\pNet\).\\
The \emph{labeling function} \(L\) labels the states with its components:
\(\forall(p,t)\in(\tr\cap\AP)\times\pl: \klab((t,p))=\{t,p\}\) and \(\forall p\in\pl: \klab(p)=\{p\}\).
The \emph{transition relation} is composed of two sets \(\krelLab=\krelLab'\cup \krelLab''\).
The relation \(\krelLab'\) connects the states with respect to the transits:
\begin{align*}
\krelLab'=&		\{(p,t,q)\in \kstates\times\tr\times \kstates\with (p,q)\in\tokenflow(t)\wedge t\not\in\AP\} \\
  \cup&		\{(p,t,(t,q))\in \kstates\times\tr\times \kstates\with (p,q)\in\tokenflow(t)\wedge t\in\AP\} \\
  \cup&		\{((t',p),t,q)\in \kstates\times\tr\times \kstates\with (p,q)\in\tokenflow(t)\wedge t\not\in\AP\} \\
  \cup&		\{((t',p),t,(t,q))\in \kstates\times\tr\times \kstates\with (p,q)\in\tokenflow(t)\wedge t\in\AP\}.
\end{align*}
The relation \(\krelLab''\) adds \(\stutterSymbol\)-labeled loops to states \(s\in\pl\)
and \(\stutterSymbol\)-labeled edges between~\((t,p)\) and \(p\) states to allow for the stuttering of finite chains:
\(\krelLab''=\{(p,\stutterSymbol,p)\in \kstates\times\{\stutterSymbol\}\times \kstates\}\cup\{((t,p),\stutterSymbol,p)\in \kstates\times\{\stutterSymbol\}\times \kstates\}\). The \emph{states} \(\kstates\) are exactly the states reachable from the initial states.

We define the function \(\ksucc:\kstates\times\katoms\to2^\kstates\)
with \(\ksucc(s,l)=\{s'\in\kstates\with (s,l,s')\in\krelLab\}\)
returning all \(l\)-labeled successors of a state \(s\).
\end{definition}

For an alternating tree automaton created of a \ctlStarText{} formula \(\ctlStarFormula\)
and a labeled Kripke structure we define an alternating word automaton
accepting the sequences of transitions which flow trees satisfy \(\ctlStarFormula\).
The definition is very similar to \cite{DBLP:journals/jacm/KupfermanVW00},
we only apply the edge definition on each equally labeled group of transitions separately.
\begin{definition}[Product Automaton]
Given an alternating tree automaton \(\tree_{\D,\ctlStarFormula}=(2^\AP,\D,\treeStates,\treeRel,\treeInit,\treeFinal)\)
accepting exactly all \(\D\)-trees satisfying \(\ctlStarFormula\) 
and a labeled Kripke structure \(\kripke_{(\pNet,\AP)}=(\AP,\katoms,\kstates,\kinit,\klab,\krelLab)\) created by \refDef{kripkeStructure}
with degrees in \(\D\).
The \emph{product automaton} \(\aba_{\neg\ctlStarFormula_i}=(\katoms,\kstates\times \treeStates,\abaRel, \kinit\times\{\treeInit\},\abaFinal)\)
of \(\tree_{\D,\ctlStarFormula}\) and \(\kripke_{(\pNet,\AP)}\) is defined with \(\abaRel\):\\
For every
\((s,q)\in\kstates\times\treeStates\),
\(l\in\katoms\),
\(\ksucc(s,l)\neq\emptyset\), and
\(\treeRel(q,\klab(s),\size{\ksucc(s,l)})=\theta\),
we have an edge \(\abaRel((s,q),l)=\theta'\),
where the positive Boolean formula \(\theta'\)
is obtained from \(\theta\)
by replacing the atoms \((c,q')\)
with \((\langle \ksucc(s,l)\rangle_c,q')\), where \(\langle \ksucc(s,l)\rangle_c\)
is the \(c\)-th value of the ordered list of successors.\\
The acceptance condition of \(\treeFinal\) is transferred to \(\abaFinal\) by preserving the type 
and building the cross product of the acceptance set(s) with~\(\kstates\).
\end{definition}

\refLemma{sizeBuchiAutomaton} states that the size of the nondeterministic B\"uchi automata is single-exponential for specifications given in \ctlText{}
and double-exponential in specifications given in \ctlStarText{}.

\begin{proof}[Size of the B\"uchi Automaton (\refLemma{sizeBuchiAutomaton})]
The construction for a \ctlStarText{} formula \(\ctlStarFormula\) and 
a set of directions \(\D\subset\N\) results 
in a hesitant alternating automaton of size \(\oclass{\size{\D}\cdot2^{\size{\ctlStarFormula}}}\)
(for \ctlText{} only of size \(\oclass{\size{\D}\cdot\size{\ctlStarFormula}}\)) \cite{DBLP:journals/jacm/KupfermanVW00}.
In our case the set of directions is
\(\D=\{0,\ldots,\mathtt{max}\{\size{\ksucc(s,l)}-1 \with s\in\kstates\wedge l\in\katoms\}\).
This is maximally
\(\D\subseteq\{0,\ldots,\mathtt{max}\{\size{\post{\tfl}{p,t}}-1 \with p\in\pl\wedge t\in\post{\pNet}{p}\}\}\),
the maximum of all numbers of successor transits of any place \(p\in\pl\).
Hence, at most \(\size{\D}\leq\size{\pl}\).
The product automaton created from the Kripke structure with 
\(\oclass{\size{\pNet}\cdot\size{\AP_i\cap\tr}+\size{\pNet}}\) states 
and \(\oclass{\size{\pNet^3}}\) edges and the hesitant alternating automaton has
\(\oclass{\size{\pl}\cdot2^{\size{\ctlStarFormula}}\cdot(\size{\pNet}\cdot\size{\AP_i\cap\tr}+\size{\pNet})}
=\oclass{2^{\size{\ctlStarFormula}}\cdot\size{\pNet}^3}\) states
and \(\oclass{(2^{\size{\ctlStarFormula}}\cdot\size{\pNet}^3)\cdot(\size{\tr}+1)}\) edges
(for CTL \(\oclass{\size{\ctlStarFormula}\cdot\size{\pNet}^3}\) states
and \(\oclass{(\size{\ctlStarFormula}\cdot\size{\pNet}^3)\cdot(\size{\tr}+1)}\) edges).
Removing the alternation results in another exponent:
the states of \(\ba_{\lnot \ctlStarFormula_i}\) are in 
\(\oclass{2^{2^{\size{\ctlStarFormula}}\cdot\size{\pNet}^3}\cdot 2^{2^{\size{\ctlStarFormula}}\cdot\size{\pNet}^3}}\)
(\(\oclass{2^{\size{\ctlStarFormula}\cdot\size{\pNet}^3}\cdot2^{\size{\ctlStarFormula}\cdot\size{\pNet}^3}}\) for \ctlText{}).
As \(\ba_{\lnot \ctlStarFormula_i}\) is a nondeterministic B\"uchi automaton, the edges of \(\ba_{\lnot \ctlStarFormula_i}\)
are in
\(\oclass{2^{2^{\size{\ctlStarFormula}}\cdot\size{\pNet}^3}\cdot\size{\tr}}\)
and in
\(\oclass{2^{\size{\ctlStarFormula}\cdot\size{\pNet}^3}\cdot\size{\tr}}\) for \ctlText{}.\hfill\qed
\end{proof}

We formally define the construction of a standard Petri net with inhibitor arcs from a Petri net with transits
and \(n\) B\"uchi automata.
\begin{definition}[Petri Net with Transits to Petri Net]
\label{def:pnwt2pn}
Given \(\pNet\) and the corresponding \(n\) B\"uchi automata \(\ba_{\lnot \ctlStarFormula_i}\).
We define the Petri net with inhibitor arcs \(\pNetMC=(\plMC,\trMC,\flMC,\inhibitorFlMC,\initMC)\) with
\begin{align*}
\plMC=\plMC_o\cup\bigcup_{i\in\{1,\ldots,n\}}\plMC_i,\qquad \trMC=\trMC_o\cup\bigcup_{i\in\{1,\ldots,n\}}\trMC_i\\
\flMC=\flMC_O\cup\flMC_C\cup\bigcup_{i\in\{1,\ldots,n\}}\flMC_i,\qquad \inhibitorFlMC={\inhibitorFlMC}_O\cup\bigcup_{i\in\{1,\ldots,n\}}{\inhibitorFlMC}_i
\end{align*}
and a partial labeling function \(\lambda:\trMC\to\tr\cup\{\stutterSymbol\}\) by:
\begin{itemize}
\item[(o)] The original part of the net is a copy of \(\pNet\) without the transits and with an additional activation place \(\act{o}\).
Furthermore, we use the places \(\normalMode\) and \(\stutteringMode\),
the switch \(t_{\normalMode\to\stutteringMode}\), and the stuttering transition \(t_\stutterSymbol\)
to allow for checking finite runs, in which case 
we have to trigger the automata \(\ba_{\lnot \ctlStarFormula_i}\) infinitely often to handle the stuttering:
\begin{align*}
\plMC_O=&\pl\cup\{\act{o},\normalMode,\stutteringMode\},\\
\trMC_O=&\tr\cup\{t_{\normalMode\to\stutteringMode},t_\stutterSymbol\},\\
\flMC_O=&\fl\cup\{(\act{o},t)\with t\in\tr\}\cup\{(\normalMode,t_{\normalMode\to\stutteringMode}),(t_{\normalMode\to\stutteringMode},\stutteringMode),(\act{o},t_\stutterSymbol)\},\\
{\inhibitorFlMC}_O=&\inhibitorFlMC\cup\{(\stutteringMode,t)\with t\in\tr\}\cup\{(\normalMode,t_\stutterSymbol)\}
\end{align*}
\item[(sub)] For each B\"uchi automaton \(\ba_{\lnot \ctlStarFormula_i}=(\tr\cup\{\stutterSymbol\},Q,Q_0,E,F) \), we create the places, transition, and flows
to simulate the automaton.
The \emph{places} are the states of the automaton with
a special place \(\initsub_i\) for initially guessing the violating tree,
two places \(\subnet{\normalMode}_i\) and \(\subnet{\stutteringMode}_i\) for switching from normal to stuttering mode,
and an activation place \(\subnet{\act{t}}_i\) for each transition \(t\in\tr\),
and one \(\subnet{\act{\stutterSymbol}}_i\) for the global stuttering transitions:
\[\plMC_i=\{\initsub_i,\subnet{\normalMode}_i,\subnet{\stutteringMode}_i,\subnet{\act{\stutterSymbol}}_i\}\cup\{\subnet{s}_i \with s\in Q\}\cup\{\subnet{\act{t}}_i\with t\in\tr\}.\]
The \emph{transitions} consists of
\[\trMC_i=\{\subnet{t_{\normalMode\to\stutteringMode}}_i\}\cup\tr_{\startfl_i}\cup\tr_{E_i}\cup\tr_{\skipT_i}\cup\tr_{\stutterSymbol_i}\]
with
the switch from normal to stuttering mode \(\subnet{t_{\normalMode\to\stutteringMode}}_i\), 
one transition for each initial flow chain:
\(\tr_{\startfl_i}=\{\subnet{t_p}_i\with t\in\tr\wedge (\startfl,p)\in\tokenflow(t)\}\),
one transition for each edge of \(\ba_{\lnot \ctlStarFormula_i}\):
\(\tr_{E_i}=\{\subnet{e}_i\with e\in E\}\),
one skipping transition \(\subnet{t_{\skipT}}_i\) for each transition \(t\in\tr\) to not trigger the automaton when
the transition does not extend the current chain:
\(\tr_{\skipT_i}=\{\subnet{t_{\skipT}}_i\with t\in\tr\}\), and
for each stuttering edge \(e=(s,\stutterSymbol,s')\) of \(\ba_{\lnot \ctlStarFormula_i}\)
there is one \emph{local} stuttering transition for each transition \(t\in\tr\) which is no label of a successor edge of the state \(s\) 
to move the activation token to the next subnet, when this net is in stuttering mode, but not the global net:
\(\tr_{\stutterSymbol_i}=\{\subnet{t_{e_{\stutterSymbol}}}_i\with e=(s,\stutterSymbol,s')\in E \wedge t\in\tr\wedge \neg\exists(s,t,s'')\in E\}\).\\
The \emph{labeling} function labels every transition \(t\MC\in\trMC\) corresponding to a transition \(t\in\tr\) or the stuttering
with \(t\) or \(\stutterSymbol\), respectively:
\(\forall\subnet{t_p}_i\in\tr_{\startfl_i}:\lambda(\subnet{t_p}_i)=t\),
\(\forall\subnet{e}_i\in\tr_{E_i}:\lambda(\subnet{e}_i)=l\) with \(e=(s,l,s')\),
\(\forall\subnet{t_{\skipT}}_i\in\tr_{\skipT_i}:\lambda(\subnet{t_{\skipT}}_i)=t\), and
\(\forall\subnet{t_{e_\stutterSymbol}}_i\in\tr_{\stutterSymbol_i}:\lambda(\subnet{t_{e_\stutterSymbol}}_i)=t\).\\
The \emph{flows} connect each transition \(t\MC\in\trMC_i\setminus\tr_{\stutterSymbol_i}\) corresponding to a transition \(t\in\tr\)
with the normal mode
and the corresponding activation token \(\subnet{\act{t}}_i\):
\begin{align*}
{\flMC_\normalMode}_i=&\{(\subnet{\act{t}}_i,t\MC)\with t\MC\in\tr_i\setminus\tr_{\stutterSymbol_i}\wedge t\in\tr \wedge \lambda(t\MC)=t\},\\
{{\inhibitorFlMC}_\normalMode}_i=&\{(\subnet{\stutteringMode}_i,t\MC)\with  t\MC\in\tr_i\setminus\tr_{\stutterSymbol_i}\wedge t\in\tr \wedge \lambda(t\MC)=t\}.
\end{align*}
The transitions for guessing a flow chain are connected to the corresponding initial state of \(\ba_{\lnot \ctlStarFormula_i}\):
\[\flMC_{\startfl_i}=\{(\initsub_i,\subnet{t_p}_i),(\subnet{t_p}_i,s_0)\with \subnet{t_p}_i\in\trMC_{\startfl_i}\wedge s_0\in Q_0\wedge (s_0=p\vee s=(t_p,p))\}.\]
The transitions corresponding to an edge of \(\ba_{\lnot \ctlStarFormula_i}\), move the tokens accordingly:
\[\flMC_{E_i}=\{(\subnet{s}_i,\subnet{e}_i),(\subnet{e}_i,\subnet{s'}_i)\with \subnet{e}_i\in\tr_{E_i} \wedge e=(s,\lambda(\subnet{e}_i),s')\}.\]
\emph{Skipping} is only allowed in situations where no corresponding transition is firable:
\[{\inhibitorFlMC}_{\skipT_i}=\{(\subnet{s}_i,\subnet{t_{\skipT}}_i)\with \subnet{t_{\skipT}}_i\in\tr_{\skipT_i}\wedge \exists (s,\lambda(\subnet{t_{\skipT}}_i),\cdot)\in E\}.\]
The \emph{stuttering} transitions are only allowed in the stuttering mode:
\({\inhibitorFlMC}_{\stutteringMode_i}=\{(\subnet{\normalMode}_i,t\MC)\with \lambda(t\MC)=\stutterSymbol\vee t\MC\in\tr_{\stutterSymbol_i}\}\),
the \emph{global} stuttering transitions move the active stuttering token
\({\flMC_{\stutteringMode_i}}^1=\{(\subnet{\act{\stutterSymbol}}_i,t\MC)\with t\MC\in\trMC_{E_i}\wedge\lambda(t\MC)=\stutterSymbol\}\),
the \emph{local} stuttering transitions move the corresponding active token
\({\flMC_{\stutteringMode_i}}^2=\{(\subnet{\act{t}}_i,t\MC)\with t\MC\in\tr_{\stutterSymbol_i}\wedge\lambda(t\MC)=t\}\)
and the state token according to the edge:
\({\flMC_{\stutteringMode_i}}^3=\{(\subnet{s}_i,\subnet{t_{e_\stutterSymbol}}_i),(\subnet{t_{e_\stutterSymbol}}_i,\subnet{s'}_i)\with
\subnet{t_{e_\stutterSymbol}}_i\in\tr_{\stutterSymbol_i}\wedge e=(s,\lambda(\subnet{e}_i),s')\}\).
With \(\flMC_{\stutteringMode_i}={\flMC_{\stutteringMode_i}}^1\cup{\flMC_{\stutteringMode_i}}^2\cup{\flMC_{\stutteringMode_i}}^3\),
the flows of the subnet are the union of the previous sets,
in addition to a nondeterministically switch from normal to stuttering mode:
\begin{align*}
\flMC_i=&{\flMC_\normalMode}_i\cup\flMC_{\startfl_i}\cup\flMC_{E_i}\cup\flMC_{\stutteringMode_i}\cup\{(\subnet{\normalMode}_i,\subnet{t_{\normalMode\to\stutteringMode}}_i),(\subnet{t_{\normalMode\to\stutteringMode}}_i,\subnet{\stutteringMode}_i)\},\\
{\inhibitorFlMC}_i=&{{\inhibitorFlMC}_\normalMode}_i\cup{\inhibitorFlMC}_{\skipT_i}\cup{\inhibitorFlMC}_{\stutteringMode_i}.
\end{align*}
\item[(con)] The nets are connected in a sequential manner:
\begin{align*}
\flMC_C=&\{(t,\subnet{\act{t}}_1)\with t\in\tr\}\cup\{(t_\stutterSymbol,\subnet{\act{\stutterSymbol}}_1)\}\\
\cup&\bigcup_{i\in\{1,\ldots,n-1\}}\{(t\MC,\subnet{\act{\stutterSymbol}}_{i+1})\with t\MC\in\trMC_i\wedge\lambda(t\MC)=\stutterSymbol \}\\
\cup&\bigcup_{i\in\{1,\ldots,n-1\}}\{(t\MC,\subnet{\act{t'}}_{i+1})\with t\MC\in\trMC_i \wedge \lambda(t\MC)=t\in\tr\}\\
\cup&\{(t\MC,\act{o})\with t\MC\in\trMC_n \wedge\lambda(t\MC)=l\in\tr\cup\{\stutterSymbol\}\}.
\end{align*}
\item[(in)] The initial marking is the original marking with the activation place for the original part and one place 
for each subnet to start guessing the chain, and one for switching into the stuttering mode:
\[\initMC=\init\cup\{\act{o},\normalMode\}\cup\{\subnet{\normalMode}_i,\initsub_i\with i\in\{1,\ldots,n\}\}.\]
\end{itemize}
\end{definition}

The size of the constructed net is dominated by the nondeterministic B\"uchi automata
\(\ba_{\lnot \ctlStarFormula_i}=(\tr\cup\{\stutterSymbol\}, Q_i, I_i, E_i,F_i)\)
checking the \ctlStarText{} subformulas.

\begin{proof}[Size of the Constructed Net (\refLemma{sizeMCNet})]
For the number of places, we have \(\size{\plMC_O}=\size{\pl}+3\) and \(\size{\plMC_i}=4+\size{Q_i}+\size{\tr}\).
Hence, \(\size{\plMC}=\size{\pl}+3+(4+\size{\tr})\cdot n+\sum_{i=1}^n\size{Q_i}\).\\
For the number of transitions, we have \(\size{\trMC_O}=\size{\tr}+2\) and
the size of  \(\trMC_i\) is in \(\oclass{1+\size{\tr}\cdot\size{\pl} + \size{E_i}+\size{\tr}+\size{Q_i}\cdot\size{\tr}}\)
because each state has a stuttering edge and maximally there is no other outgoing transition.
Hence, \(\size{\trMC}\) is in \(\oclass{\size{\tr}+2+(1+\size{\tr}\cdot\size{\pl} + \size{\tr})\cdot n+\sum_{i=1}^n\size{E_i}+\tr\cdot\sum_{i=1}^n\size{Q_i}}\).
For a double-exponential number of states \(Q_i\) and edges \(E_i\) for specifications \(\ctlStarFormula_i\) in \ctlStarText{} and
single-exponential for specifications in \ctlText{} (\refLemma{sizeBuchiAutomaton})
the size of the constructed net is in the respective classes.
\hfill\qed
\end{proof}

The size of the formula is dependent on the size of the net and therewith also dominated by 
the nondeterministic B\"uchi automaton construction.

\begin{proof}[Size of the Constructed Formula (\refLemma{sizeFormula})]
The next and the until replacement introduces disjunctions over all transitions 
of the net. Also, the skipping constraint uses nearly every transition.
Hence, the size of the formula depends on the number of transitions \(\trMC\) and
is therewith double-exponential for specifications in \ctlStarText{} and single-exponential
for specifications in \ctlText{}.
\hfill\qed
\end{proof}

The correctness of the transformation is based on the correctness of the Kripke structure and the automata constructions.
The constructed formula and the constructed net together ensure
the correct triggering of the automata. 
\begin{proof}[Correctness of the Transformation (\refLemma{correctnessTransformation})]
The unwinding of the Kripke structure along each transition sequence of a run
creates trees corresponding to flow trees of the run.
The standard constructions yield the correspondence of the acceptance
of the automaton which we check with \(\lnot\ltlAlways\ltlEventually\bigvee_{b\in F} \subnet{b}_i\).
Because of the ``infinitely often'', we do not have to do anything special for the net structure.
By checking all runs of the net \(\pNetMC\) and the nondeterministic guessing of the violating
flow tree in each subnet we check all flow trees.
The run part of the formula is adequately substituted such that all elements
concerning the timeline (the \(\ltlUntil\) and the \(\ltlNext\) operator) regarding \(\pNet\)
are adapted such that the sequential steps are omitted regarding the timeline of \(\pNetMC\).

The open part of this construction is to check whether only correct runs
regarding the sequential triggering and stuttering of the subnets are allowed.\\
\emph{All nets in normal mode:} The original part \(\pNetMC_O\) can only 
choose one transition~\(t\in\tr\) and move the active token to the first subnet.
The stuttering transition \(t_\stutterSymbol\) is not enabled due to the inhibitor arc to \(\normalMode\).
If in the current state of the subnet \(t\) can extend the tree, i.e.,
is the label of a successor arc of the state, the only transition fireable (apart from the switch)
is the corresponding transition in \(\tr_{E_i}\) (because of the inhibitor arcs for \(\tr_{\skipT_i}\)).
If not, only the corresponding skipping transition in \(\tr_{\skipT_i}\) is fireable and moves the token to the next subnet.
If the net wrongly chose to be in normal mode, i.e., this subnet could only fire skipping transitions,
the constraint 
\(\neg\ltlEventually\ltlAlways((\bigvee_{t\in\trMC_i} t)\rightarrow(\bigvee_{t'\in\tr_{\skipT_i}} t'))\)
of the formula omits these runs.\\
\emph{Global stuttering:} If at some point the global net switches into the stuttering mode, only transition \(t_\stutterSymbol\)
is fireable anymore. This transition only activates the first net with the stutter token in \(\subnet{\act{\stutterSymbol}}_1\).
Since all but the global stuttering transitions depend on an active place corresponding to a transition
only the global stuttering transitions \(\subnet{e}_i\in\tr_{E_i}\) with \(e=(s,\stutterSymbol,s')\) are fireable.
Hence, when the net does not switch into the stuttering mode, the whole run is stuck and is not considered due to the 
\((\ltlAlways\ltlEventually\act{o})\) constraint in the formula \(\phiRun\MC\).\\
\emph{Local stuttering:} If the global net is not in stuttering mode \(\stutteringMode\),
but some net \(\pNetMC_i\) is, then only the transitions in \(\tr_{\stutterSymbol_i}\)
are fireable. If the net wrongly chose to be in stuttering mode then it would stuck and again the constraint 
\((\ltlAlways\ltlEventually\act{o})\) of the formula omits this run.

The run which never decides to track any tree for a subnet introduces no problems because the initial place \(\initsub_i\)
cannot be part of the B\"uchi places.
\hfill\qed
\end{proof}

The previous lemmata yield the final complexity results.
\begin{proof}[\refTheo{complexityMC}]
For a Petri net with transits \(\pNet\) and a Flow-\ctlStarText{} formula~\(\phiRun\),
\refLemma{sizeMCNet} and \refLemma{sizeFormula} yield the double-exponential size
of the constructed Petri net with inhibitor arcs \(\pNetMC\)
and the constructed formula \(\phiLTL\MC\) (the single-exponential size for the \ctlText{} fragment).
\refLemma{correctnessTransformation} yields the correctness.
Checking a safe Petri net against LTL can be seen as checking a Kripke structure of exponential size (due to the markings of the net).
Since this can be checked in linear time in the size of
the state space and in exponential time in the size of the formula \cite{DBLP:reference/mc/2018},
we obtain a triple-exponential algorithm for \ctlStarText{} formulas 
and a double-exponential algorithm for \ctlText{} formulas in the size of the net and the formula.
\hfill\qed
\end{proof}

\end{document}